\documentclass{article}

\usepackage{arxiv}

\usepackage[utf8]{inputenc} 
\usepackage[T1]{fontenc}    
\usepackage{hyperref}       
\usepackage{url}            
\usepackage{booktabs}       
\usepackage{amsfonts}       
\usepackage{nicefrac}       
\usepackage{microtype}      
\usepackage{graphicx}
\usepackage{caption}
\usepackage{subcaption}
\usepackage[export]{adjustbox}
\usepackage{amsmath}
\usepackage{algorithm}
\usepackage{algorithmic}
\usepackage{mathtools}
\usepackage{doi}
\usepackage{float}
\usepackage{tikz} 
\usetikzlibrary{bayesnet} 

\title{Outcome-guided spike-and-slab Lasso Biclustering: A Novel Approach for Enhancing Biclustering Techniques for Gene Expression Analysis}

\author{ \href{https://orcid.org/0000-0001-8462-9970}{\includegraphics[scale=0.06]{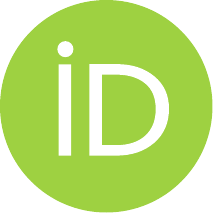}\hspace{1mm}Luis A.~Vargas-Mieles}\\
	CITIID\thanks{Cambridge Institute for Therapeutic Immunology and Infectious Disease.}\\
	University of Cambridge\\
	Cambridge, UK \\
	\texttt{lv375@cam.ac.uk} \\
        \And
	\href{https://orcid.org/0000-0002-5931-7489}{\includegraphics[scale=0.06]{orcid.pdf}\hspace{1mm}Paul D. W.~Kirk}\\
	CITIID\footnotemark[1]\\
        MRC Biostatistics Unit\\
	University of Cambridge\\
	Cambridge, UK \\
	\texttt{paul.kirk@mrc-bsu.cam.ac.uk} \\
	\And
	\href{https://orcid.org/0000-0001-9755-1703}{\includegraphics[scale=0.06]{orcid.pdf}\hspace{1mm}Chris~Wallace}\\
	CITIID\footnotemark[1]\\
        MRC Biostatistics Unit\\
	University of Cambridge\\
	Cambridge, UK \\
	\texttt{cew54@cam.ac.uk} \\
}

\renewcommand{\shorttitle}{Outcome-guided Spike-and-Slab Lasso Biclustering}
\DeclareMathOperator*{\argmax}{arg\,max}

\hypersetup{
pdftitle={Outcome-guided Spike-and-Slab Lasso Biclustering},
pdfsubject={q-bio.NC, q-bio.QM},
pdfauthor={Luis A.~Vargas-Mieles, Paul D. W.~Kirk, Chris~Wallace},
pdfkeywords={Biclustering, Factor analysis, Profile regression, Spike-and-Slab Lasso},
}

\begin{document}
\maketitle

\begin{abstract}
    Biclustering has gained interest in gene expression data analysis due to its ability to identify groups of samples that exhibit similar behaviour in specific subsets of genes (or vice versa), in contrast to traditional clustering methods that classify samples based on all genes. Despite advances, biclustering remains a challenging problem, even with cutting-edge methodologies. This paper introduces an extension of the recently proposed Spike-and-Slab Lasso Biclustering (SSLB) algorithm, termed Outcome-Guided SSLB (OG-SSLB), aimed at enhancing the identification of biclusters in gene expression analysis. Our proposed approach integrates disease outcomes into the biclustering framework through Bayesian profile regression. By leveraging additional clinical information, OG-SSLB improves the interpretability and relevance of the resulting biclusters. Comprehensive simulations and numerical experiments demonstrate that OG-SSLB achieves superior performance, with improved accuracy in estimating the number of clusters and higher consensus scores compared to the original SSLB method. Furthermore, OG-SSLB effectively identifies meaningful patterns and associations between gene expression profiles and disease states. These promising results demonstrate the effectiveness of OG-SSLB in advancing biclustering techniques, providing a powerful tool for uncovering biologically relevant insights. The OGSSLB software can be found as an R/C++ package at \url{https://github.com/luisvargasmieles/OGSSLB}.
\end{abstract}


\section{Introduction}
Over the last few decades, the identification of groups that share interesting common characteristics has been a key objective in various real-world applications. Clustering has proven to be a crucial method for discovering these groups that exhibit patterns within high-dimensional data, particularly in the context of large omics datasets \cite{Zhao2005}. This technique enables the detection of associations between related entities based on shared features or attributes.

One domain of omics in which clustering techniques have been widely employed has been the examination of transcriptomic data, which captures patterns of gene expression levels within biological entities such as tissues or cells \cite{1339264}. The need to understand shared transcriptional patterns embedded in gene expression data has led to extensive development of clustering methodologies.

Although clustering has been beneficial in revealing hidden patterns within these large-scale datasets, it is not without drawbacks. Among its disadvantages, traditional clustering models assume that samples within a cluster behave similarly across all genes and vice versa. Additionally, clustering often results in a partition of the samples or genes into disjoint subsets. These assumptions may oversimplify the biological system under analysis.

Owing to these limitations, biclustering, a methodology that clusters genes and samples simultaneously, has gained more attention in recent years \cite{10.1093/bioinformatics/btl060}. This approach allows flexibility in capturing subsets of genes that may behave differently across conditions or subsets of samples that differ according to specific sets of features. Furthermore, biclustering permits overlapping patterns, where genes or samples may belong to multiple biclusters, which more closely matches biological systems. For instance, samples may cluster by sex, disease age and disease state simultaneously, and a single gene may be a member of two or more biological pathways.

Several approaches have been proposed to estimate these subgroups of genes and samples, and various reviews of the biclustering methods developed over the past decades exist in the literature (see, e.g., \cite{1324618, 10.1093/bib/bbs032, padilha2017systematic}). Building upon the results of \cite{10.1093/bib/bbab140}, this work focuses on the adoption of a multiplicative model, which has proven advantageous in this context. Such models have demonstrated effective capture of diverse sources of variability in gene expression data, including the presence of outlier genes or genes with fluctuating expression levels \cite{10.1093/bioinformatics/btq227}.

Among the existing algorithms using this methodology, we highlight three: Factor analysis for bicluster acquisition (FABIA) \cite{10.1093/bioinformatics/btq227}, the BicMix biclustering method \cite{10.1371/journal.pcbi.1004791} and Spike-and-Slab Lasso Biclustering (SSLB) \cite{10.1214/20-AOAS1385}. All three are based on a Bayesian factor analysis model with sparsity-inducing priors that have proven to possess a notable ability to recover latent structures in gene expression data. However, a comparative study by \cite{10.1093/bib/bbab140} highlights that SSLB has the advantage of allowing different sparsity levels on each bicluster, in comparison to BicMix, which allows only two levels of sparsity (sparse or dense) for each bicluster, and FABIA, which uses the same sparsity level for all biclusters. Furthermore, while FABIA requires setting the number of biclusters in advance, SSLB (and BicMix) automatically estimates the number of biclusters.

Although recent approaches, such as the ones mentioned above, have been shown to be capable of revealing these latent structures within gene expression data, biclustering, in general, is recognised as an NP-hard problem \cite{10.1093/bioinformatics/18.suppl_1.S136, PEETERS2003651}. The NP-hardness arises from the challenge of simultaneously grouping rows and columns of a matrix to identify coherent submatrices while considering various constraints and optimisation criteria. Added to the fact that biclusters can also overlap, these difficulties pose a substantial challenge to even the most state-of-the-art methods, further complicating the accurate identification of samples and gene groups that share a common characteristic.

One promising approach to mitigate this complexity is the integration of informative outcome data into the clustering process, thereby guiding the inference towards biologically relevant clustering structures. Several outcome-guided clustering methods have been developed in recent years, with applications in K-Means clustering \cite{10.1111/rssc.12536} and gene selection based on survival data \cite{10.1093/bioinformatics/btq470}, to mention a few. For additional insights, see \cite{10.1371/journal.pbio.0020108, https://doi.org/10.1002/wics.1270}.

In light of these developments, Bayesian profile regression has emerged as another outcome-guided, semi-supervised method for clustering that leverages an outcome variable to inform cluster allocations \cite{10.1093/biostatistics/kxq013, JSSv064i07}. Unlike some of the previously mentioned approaches, it offers a fully model-based framework that can handle a variety of outcome types, making it more versatile. This approach has already shown success in handling binary covariate data \cite{https://doi.org/10.1002/sim.10119} as well as longitudinal or multivariate continuous outcomes \cite{10.1093/jrsssc/qlad097}, making it a valuable tool in the context of gene expression analysis.

Building on this success, we explore whether such outcome-guided strategies can also enhance biclustering, where the goal is to simultaneously group genes and samples. Since most gene expression studies also include phenotype information such as age, sex, and disease status, we investigate in this work whether integrating disease outcomes would enhance cluster consistency within specific disease groups. To achieve this, we introduce an outcome-guided version of SSLB by incorporating the disease outcome of the samples into the model via Bayesian profile regression, aiming to better guide the biclustering membership of genes and samples. As we show in numerical and real-data experiments, this integration enhances the accuracy of the SSLB model and refines the biological relevance of the estimated biclusters.

The remainder of the paper is organised as follows. Section \ref{sec:problem_statement} discusses the current state of biclustering models, particularly within the context of factor analysis models, and explains the SSLB model that we aim to improve. Section \ref{sec:methodology} highlights the potential contributions of additional data available in most gene expression studies and presents the methodology used in our work, which integrates Bayesian profile regression into the SSLB model, detailing the computations added to implement this new approach. Section \ref{sec:numerical_results} details the results of our experiments and provides a comprehensive analysis of the findings. Finally, Section \ref{sec:conclusions} concludes by summarising the key contributions of our research and suggesting directions for future work.

\section{Factor Analysis Models and Current Biclustering Techniques}
\label{sec:problem_statement}
We assume that gene expression data is represented in a matrix $\mathbf{X} \in \mathbb{R}^{N \times G}$ with $N$ samples and $G$ features or genes. Additionally, we assume that the data $\mathbf{X}$ contains $K$ non-disjoint latent groups of genes and samples that potentially may be linked due to some common biological characteristics. The problem of determining the number of subgroups $K$, as well as identifying the genes and samples belonging to each of these groups, is defined as ``biclustering".

Following the results of \cite{10.1214/20-AOAS1385}, we adopt a factor analysis model to identify these latent groups or biclusters, where $\mathbf{X}$ can be represented as
\begin{equation}
\label{eqn:mult_model_general}
    \mathbf{X} = \mathbf{\Lambda Z^{\mathsf{T}}} + \mathbf{E},
\end{equation}
where
\begin{itemize}
    \item $\mathbf{\Lambda} \in \mathbb{R}^{N \times K}$, which will be called the sample loading matrix,
    \item $\mathbf{Z} \in \mathbb{R}^{G \times K}$, which will be called the gene loading matrix, and
    \item $\mathbf{E} = \left[\varepsilon_1, \ldots, \varepsilon_N\right]^{\mathsf{T}} \in \mathbb{R}^{N \times G}$ is noise, where each $\varepsilon_i \sim N_G(\mathbf{0}, \Sigma)$, and $\Sigma = \mathrm{diag} \{\sigma_j^2\}_{j=1}^G$.
\end{itemize}

Current biclustering algorithms that use the model given in \eqref{eqn:mult_model_general} implement an unsupervised approach to identify sparse groups of relevant genes and samples and thus infer $\boldsymbol{\Lambda}$ and $\mathbf{Z}$. Their main input is $\mathbf{X}$ without additional information (apart from the necessary model parameters) included in the model. For instance, FABIA \cite{10.1093/bioinformatics/btq227} uses a Laplacian prior on all $\boldsymbol{\Lambda}$ and $\mathbf{Z}$ entries to induce sparsity, applying the same prior to every entry in these matrices. BicMix \cite{10.1371/journal.pcbi.1004791}, on the other hand, allows the columns of $\boldsymbol{\Lambda}$ and $\mathbf{Z}$ to be sparse or dense. For the sparse components, it utilises three levels of shrinkage, each employing a three-parameter beta (TPB) prior \cite{NIPS2011_ad972f10}, to promote sparsity.

Finally, SSLB employs the Spike-and-Slab Lasso prior \cite{doi:10.1080/01621459.2016.1260469} for both $\boldsymbol{\Lambda}$ and $\mathbf{Z}$. This prior enables stronger regularisation on near-zero coefficients (in the spike) to achieve sparsity, while applying weaker regularisation on larger coefficients (in the slab) to maintain accuracy. A key advantage of the SSLB prior over FABIA and BicMix is its ability to allow varying levels of sparsity for each bicluster. As mentioned earlier, FABIA uses the same prior for all biclusters, and BicMix permits only two sparsity levels (‘sparse’ or ‘dense’). SSLB, however, assigns a distinct sparsity parameter to each bicluster.

\subsection{The SSLB model}

For completeness, we provide a brief explanation of each component within this Bayesian model. See \cite{10.1214/20-AOAS1385} for more details.

\subsubsection{SSLB likelihood}
Since \eqref{eqn:mult_model_general} can also be written as
\begin{equation*}
    \mathbf{X}=\sum_{k=1}^K \boldsymbol{\lambda}^k \mathbf{z}^{k \mathsf{T}}+\mathbf{E},
\end{equation*}
where the superscript $\boldsymbol{\lambda}^k$ represents the $k$th column of $\boldsymbol{\Lambda}$, the likelihood is defined as

\begin{equation*}
    p (\mathbf{X} \mid \boldsymbol{\Lambda}, \mathbf{Z}) \propto \prod_{i=1}^N\left\{\exp \left[-0.5\left(\mathbf{x}_i - \boldsymbol{\lambda}_i \mathbf{Z}^{\mathsf{T}} \right)^{\mathsf{T}} \boldsymbol{\Sigma}^{-1}\left(\mathbf{x}_i - \boldsymbol{\lambda}_i \mathbf{Z}^{\mathsf{T}} \right)\right]\left(\prod_{j=1}^G \sigma_j^2\right)^{-1 / 2}\right\},
\end{equation*}
where the subscript $\boldsymbol{\lambda}_i$ refers to the $i$th row of $\boldsymbol{\Lambda}$\footnote{Throughout, this superscript and subscript notation is utilised to denote the respective column and row vector of a matrix.}.

\subsubsection{SSLB priors}
\label{subsubsec:sslb_priors}
\paragraph{Prior on the elements of \texorpdfstring{$\mathbf{Z}$}{Z}}
For the elements of the gene loading matrix, we have a spike-and-slab prior \cite{doi:10.1080/01621459.2016.1260469}, defined by
\begin{equation*}
    p (\mathbf{Z} \mid \boldsymbol{\Gamma}, \omega_0, \omega_1) \propto \prod_{j=1}^G \prod_{k=1}^K \left[\left(1-\gamma_{j k}\right) \omega_0 \exp \left(-\omega_0\left|z_{j k}\right|\right)+\gamma_{j k} \omega_1 \exp \left(-\omega_1\left|z_{j k}\right|\right)\right],
\end{equation*}
where $\boldsymbol{\Gamma} = \{ \gamma_{jk} \}_{j, k =1}^{G, K}$ are binary indicator variables that specify if feature $j$ is active in bicluster $k$. Depending on $\gamma_{jk}$, each $z_{jk}$ can be drawn from either a Laplacian ``spike" characterised by a large parameter value $\omega_0$ and is consequently negligible, or from a Laplacian ``slab" with a small parameter $\omega_1$ and, consequently, can be large. Refer to Section \ref{subsec:impl_sslb_initial_cond} for detailed information on the values of $\omega_0$ and $\omega_1$.

\paragraph{Prior on the gene binary indicator variable \texorpdfstring{$\boldsymbol{\Gamma}$}{Γ}}
To estimate each $\{ \gamma_{jk} \}_{j, k =1}^{G, K}$, the authors use the Beta-Bernoulli prior

\begin{equation*}
    p ( \boldsymbol{\Gamma} \mid \boldsymbol{\Theta}, \alpha) \propto \prod_{j=1}^G \prod_{k=1}^{K} \theta_k^{\gamma_{j k}+\alpha-1}\left(1-\theta_k\right)^{1-\gamma_{j k}},
\end{equation*}
where
\begin{itemize}
    \item $\boldsymbol{\Theta} = \{\theta_1, \ldots, \theta_K \}$,
    \item $\gamma_{j k} \mid \theta_k \sim \mathrm{Bernoulli}(\theta_k)$,
    \item $\theta_k \sim \mathrm{Beta}(\alpha, 1)$.
\end{itemize}

For this prior, \cite{10.1214/20-AOAS1385} recommends a finite approximation of the Indian buffet process (IBP) prior using $\alpha = 1 / K$. When $K \rightarrow \infty$, this prior is the IBP prior. See \cite{NIPS2005_2ef35a8b} for details.

\paragraph{Prior on the elements of \texorpdfstring{$\boldsymbol{\Lambda}$}{Λ}}
For the elements of the sample loading matrix, the authors proposed an alternate formulation of the Spike-and-Slab Lasso prior previously defined for the gene loading matrix, for computational purposes. Firstly, an auxiliary variable $\{\tau_{ik}\}_{i,k = 1}^{N, K}$ is introduced in the model, such as

\begin{equation}
\label{eqn:prob_lambda_given_tau}
    \lambda_{i k} \mid \tau_{i k} \sim N(0, \tau_{i k}),
\end{equation}
and then, for each $\tau_{ik}$, a mixture of exponentials is defined as
\begin{equation} \label{eqn:prob_tau_given_gamma_tilde}
    p ( \mathbf{T} \mid \tilde{\boldsymbol{\Gamma}}, \tilde{\omega}_0, \tilde{\omega}_1) \propto \prod_{i=1}^N \prod_{k=1}^{K} \Bigg[ \left(1-\tilde{\gamma}_{i k}\right) \tilde{\omega}_0 \exp \left(-0.5 \tilde{\omega}_0 \tau_{i k}\right) + \tilde{\gamma}_{i k} \tilde{\omega}_1 \exp \left(-0.5 \tilde{\omega}_1 \tau_{i k}\right) \Bigg],
\end{equation}
where $\tilde{\boldsymbol{\Gamma}}=\left\{\tilde{\gamma}_{i k}\right\}_{i, k=1}^{N, K}$ are binary indicator variables indicating bicluster membership on the elements of $\lambda_{i k}$, and $\mathbf{T}=\left\{\tau_{i k}\right\}_{i, k=1}^{N, K}$ are the covariances of $\lambda_{i k}$. In summary, the authors represent the Laplace distribution as a scale mixture of a normal with an exponential mixing density: a spike-and-slab Lasso prior on each $\lambda_{i k}$ by introducing auxiliary variables $\tau_{i k}$ for the variance of every $\lambda_{i k}$, and then each $\tau_{i k}$ is assigned a mixture of exponentials (spike-and-slab) priors. Marginalising over the $\tau_{i k}$ yields the usual spike-and-slab Lasso prior.

\paragraph{Prior on the sample indicator variable \texorpdfstring{$\tilde{\boldsymbol{\Gamma}}$}{Gamma-tilde}}
For this variable, the authors proposed an Indian Buffet Process (IBP) prior with an optional Pitman-Yor (PY) extension prior \cite{pmlr-v2-teh07a}, defined as
\begin{align}
    \tilde{\gamma}_{i k} &\sim \mathrm{Bernoulli}(\tilde{\theta}_{(k)}) \nonumber \\
    \tilde{\theta}_{(k)} &= \prod_{l=1}^k \nu_{(l)} \nonumber \\
    \nu_{(l)} & \sim \mathrm{Beta}(\tilde{\alpha} + ld, 1 - d), \; \; \text{where} \; d \in [0, 1), \; \tilde{\alpha} > - d. \label{eqn:prob_gamma_tilde_given_theta_tilde}
\end{align}

When $0 < d < 1$, the above formulation corresponds to the Pitman-Yor IBP prior. In the case where $d = 0$, it represents the standard IBP prior. For the simulations carried out in the SSLB paper and for consistency in this work, the finite approximation to the IBP is also used for comparison, which involves a Beta prior on the sparsity weights, $\tilde{\theta}_k \sim \text{Beta}(\tilde{a}, \tilde{b})$ where $\tilde{a} \propto 1 / K$ and $\tilde{b} = 1$. See \cite{pmlr-v2-teh07a} for further details.

\paragraph{Prior on the covariance matrix \texorpdfstring{$\boldsymbol{\Sigma}$}{Σ} of \texorpdfstring{$\varepsilon_i$}{εi}}
For the covariance, $\boldsymbol{\Sigma}$, of the vectors $\varepsilon_i$ that define $\mathbf{E}$ in \eqref{eqn:mult_model_general}, an inverse gamma prior was assumed. That is
\begin{equation*}
    p (\boldsymbol{\Sigma} \mid \eta, \xi) \propto \prod_{j=1}^G\left[\left(\sigma_j^2\right)^{-(\eta / 2+1)} \exp \left(\frac{-\eta \xi}{2 \sigma_j^2} \right)\right],
\end{equation*}
where the SSLB authors suggest setting $\eta = 3$ and choosing $\xi$ such that the $95\%$ quantile of the prior on $\{\sigma_j^2\}_{j=1}^G$ matches the sample column variance $\{s_j^2\}_{j=1}^G$, i.e., $p(\sigma_j < s_j) = 0.95$. Refer to \cite[Section 2.2.4]{10.1214/09-AOAS285} and \cite[Section 2.5]{10.1214/20-AOAS1385} for further information.

\begin{figure}[h!]
    \centering
    \begin{tikzpicture}

    \node[obs]  (X)  {$x_{ij}$};
    \node[latent, above=of X, xshift=-2cm] (Lambda) {$\lambda_{ik}$};
    \node[latent, above=of X, xshift=2cm] (Z) {$z_{jk}$};
    \node[latent, below=of X, yshift=-1cm] (E) {$e_{ij}$};
    \node[latent, right=of E, yshift=-1cm] (Sigma) {$\sigma_{j}$};

    \node[latent, right=of Z, xshift=1cm] (Gamma) {$\gamma_{jk}$};
    \node[latent, left=of Gamma, yshift=1.5cm] (Theta) {$\theta_k$};
    \node[const, above=of Theta, yshift=1cm] (alpha) {$\alpha$};

    \node[latent, left=of Lambda, xshift=-1cm] (Tau) {$\tau_{ik}$};
    \node[latent, right=of Tau, yshift=1.5cm] (Gammatilde) {$\tilde{\gamma}_{ik}$};
    \node[latent, right=of Gammatilde, yshift=1.5cm] (Thetatilde) {$\tilde{\theta}_{(k)}$};
    \node[latent, above=of Thetatilde] (nu) {$\nu_{(l)}$};
    \node[const, above=of nu] (alpha_tilde) {$\tilde{\alpha}$};
    \node[const, right=of alpha_tilde, xshift=0.5cm] (d) {$d$};

    \node[const, below=of Sigma, yshift=-0.5cm] (eta) {$\eta$};
    \node[const, right=of eta, xshift=0.5cm] (xi) {$\xi$};

    \edge {Lambda,Z,E} {X}; 
    \edge {Sigma} {E}; 
    \edge {eta, xi} {Sigma}; 
    \edge {Gamma} {Z}; 
    \edge {Theta} {Gamma}; 
    \edge {alpha} {Theta}; 
    \edge {Tau} {Lambda}; 
    \edge {Gammatilde} {Tau}; 
    \edge {Thetatilde} {Gammatilde}; 
    \edge {nu} {Thetatilde}; 
    \edge {alpha_tilde,d} {nu}; 

    \plate {plate1} {(X)(E)(Lambda)(Tau)(Gammatilde)} {$i = 1, \dots, N$}; 
    \plate {plate2} {(X)(E)(Sigma)(Z)(Gamma)} {$j = 1, \dots, G$}; 
    \plate {plate3} {(Lambda)(Z)(Gamma)(Tau)(Gammatilde)(Thetatilde)} {$k = 1, \dots, K$}; 
    \plate {plate4} {(nu)} {$l = 1, \dots, k$}; 

    \end{tikzpicture}
    \caption{DAG for the SSLB-IBP model, where the indices $i = 1, \dots, N$ correspond to the $N$ samples, the indices $j = 1, \dots, G$ correspond to the $G$ genes, and the indices $k = 1, \dots, K$ represent the $K$ biclusters. Variables $x_{ij}$ are observed and correspond to the gene expression data.}

    \label{fig:dag_sslb}
\end{figure}
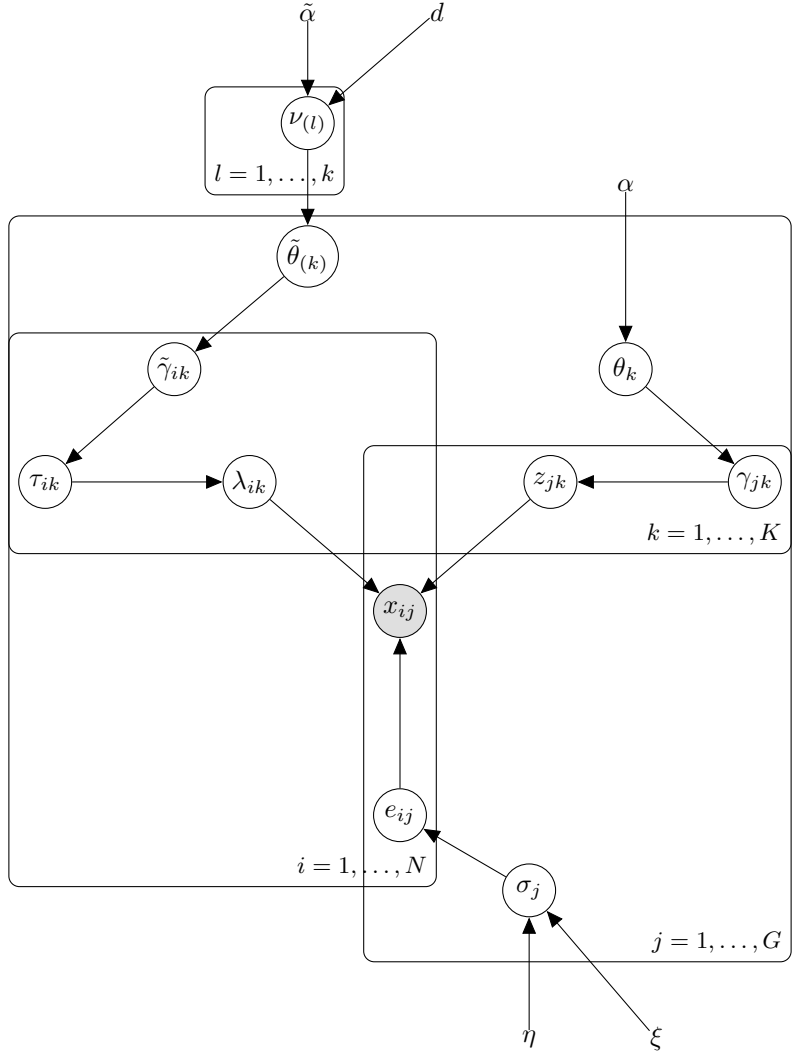

After explaining the whole hierarchical structure of the SSLB model, we translate all this information into a Directed Acyclic Graph (DAG) for a visual representation of the variables and their dependencies. See Figure \ref{fig:dag_sslb}.

\subsubsection{Estimation of biclusters: EM algorithm}
To proceed with the estimation of the parameters of interest in the SSLB model, the authors implemented an Expectation Maximisation (EM) algorithm. For completeness, we are going to briefly describe the important parts of this procedure for the IBP prior case for $\tilde{\boldsymbol{\Gamma}}$. See \cite[Section 2.3]{10.1214/20-AOAS1385} and its supplementary material for more details.

\paragraph{E step}
At iteration $t+1$ of the EM algorithm, the E step involves the computation of the following expectation
\begin{equation*}
    Q(\boldsymbol{\Delta} \mid \boldsymbol{\Delta}^{(t)})=\mathbb{E}_{\mathbf{\Lambda}, \widetilde{\Gamma} \mid \Delta^{(t)}, \mathbf{X}}[\log p (\boldsymbol{\Delta}, \mathbf{\Lambda}, \widetilde{\boldsymbol{\Gamma}} \mid \mathbf{X})],
\end{equation*}
where $\boldsymbol{\Delta}=\{\mathbf{Z}, \boldsymbol{\Sigma}, \mathbf{T}, \nu\}$ are the variables at which $Q(\boldsymbol{\Delta} \mid \boldsymbol{\Delta}^{(t)})$ will be maximised in the M step. See \eqref{eqn:log_post_no_prof_reg} and \eqref{eqn:exp_log_post_no_prof_reg} for more details.

\paragraph{M step}
In this stage, the following is calculated
\begin{equation*}
    \boldsymbol{\Delta}^{(t + 1)} = \operatorname*{arg\,max}_{\Delta} Q(\boldsymbol{\Delta} \mid \boldsymbol{\Delta}^{(t)}).
\end{equation*}

\subsubsection{Implementation of SSLB \& initial conditions}
\label{subsec:impl_sslb_initial_cond}
The SSLB algorithm employs the previously described EM algorithm combined with a dynamic posterior exploration approach to estimate $\mathbf{Z}$. This involves a gradual increase of the spike parameter $\omega_0$ through a sequence of values, while the slab parameter $\omega_1$ is kept fixed. This strategy helps stabilise large coefficients and progressively thresholds negligible coefficients to zero (see \cite[Section 2.3]{10.1214/20-AOAS1385} for more details).

The SSLB algorithm is initialised with entries of $\mathbf{Z}$ generated independently from a standard normal distribution. The entries of $\mathbf{T}$, the matrix of auxiliary variance parameters, are set to 100, representing an initial relatively non-informative prior to $\boldsymbol{\Lambda}$. The sparsity weights, $\theta_k$, are initialised at 0.5. The IBP parameters, $\boldsymbol{\nu}$, are generated independently from a $\mathrm{Beta}(1, 1)$ distribution and then ordered from largest to smallest. In real-world applications, the recommended initialisation for K, the number of biclusters, is set to $K_{init} = 50$. See \cite[Section 2.5]{10.1214/20-AOAS1385} for more details.

\section{Methodology}
\label{sec:methodology}
Considering the potential availability of additional data on sampled individuals, such as disease status, our objective is to investigate whether incorporating this information can enhance the complex task of biclustering. Notably, to the best of our knowledge, no existing biclustering method has explored the inclusion of the disease status of samples within the model.

We propose incorporating an outcome variable $\mathbf{Y} \in \{0, 1\}^{N \times C}$ into the current SSLB model, which will correspond to the presence or absence (i.e., $1$ or $0$, respectively) of disease $c \in \{1, \ldots, C\}$ in the sample $i \in \{1, \ldots, N\}$.

We assume we have gene expression data $\mathbf{X}$ also modelled as \eqref{eqn:mult_model_general}, and outcomes $\mathbf{Y}$. Since $\mathbf{Y}$ provides only sample-wise information, it will only affect the distribution related to the sample loading matrix $\boldsymbol{\Lambda}$. The general model considered in \cite{10.1093/biostatistics/kxq013} and adapted for the biclustering problem with respect to $\boldsymbol{\Lambda}$ is defined as:

\[ p \left(\boldsymbol{\Lambda} , \mathbf{Y} \mid \boldsymbol{\Theta}_{\boldsymbol{\Lambda}} , \boldsymbol{\Theta}_{\mathbf{Y}} \right) = \prod_{i=1}^N p (\mathbf{y}_i \mid  \boldsymbol{\theta}_{\mathbf{y}_i}, \boldsymbol{\lambda}_i) p (\boldsymbol{\lambda_i} \mid \boldsymbol{\theta}_{\boldsymbol{\lambda}_i} ) , \] where $\boldsymbol{\theta}_{\boldsymbol{\lambda}_i}$ represents the parameters of the model for $\boldsymbol{\lambda}_i$, and $\boldsymbol{\theta}_{\mathbf{y}_i}$ represents the parameters of the model for $\mathbf{y}_i$. Furthermore, in the profile regression setting, the factor loadings \( \boldsymbol{\Lambda} \) and the outcome \( \mathbf{Y} \) are conditionally independent because their relationship is mediated by the binary indicator matrix \( \tilde{\boldsymbol{\Gamma}} \), which governs the structure of the sampling clustering. \( \tilde{\boldsymbol{\Gamma}} \) determines which latent factors contribute to \( \boldsymbol{\Lambda} \) and how they align with \( \mathbf{Y} \). Therefore, the profile regression model becomes:

\[ p \left(\boldsymbol{\Lambda} , \mathbf{Y} \mid \boldsymbol{\Theta}_{\boldsymbol{\Lambda}} , \boldsymbol{\Theta}_{\mathbf{Y}} \right) = \prod_{i=1}^N p (\mathbf{y}_i \mid  \boldsymbol{\theta}_{\mathbf{y}_i}) p (\boldsymbol{\lambda_i} \mid \boldsymbol{\theta}_{\boldsymbol{\lambda}_i} ) , \]
where

\begin{itemize}
    \item $\boldsymbol{\Theta}_{\boldsymbol{\Lambda}} = \{ \tilde{\boldsymbol{\Gamma}}, \mathbf{T},  \nu,  \widetilde{\omega_0}, \widetilde{\omega_1}, \tilde{\boldsymbol{\theta}}, \tilde{\alpha} \}$.
    \item $p \left(\mathbf{\Lambda} \mid \boldsymbol{\Theta}_{\boldsymbol{\Lambda}}\right) \propto p ( \mathbf{\Lambda} \mid \mathbf{T}) p ( \mathbf{T} \mid \tilde{\boldsymbol{\Gamma}}, \widetilde{\omega_0}, \widetilde{\omega_1}) p(\mathbf{\tilde{\Gamma}} \mid \tilde{\boldsymbol{\theta}}, \tilde{\alpha})$.
\end{itemize}
and $\boldsymbol{\Theta}_{\mathbf{Y}}$ is the set of bicluster-specific parameters of the model for $\mathbf{Y}$, which also includes \( \tilde{\boldsymbol{\Gamma}} \). Note that each of the probability density functions given in $p \left(\mathbf{\Lambda} \mid \boldsymbol{\Theta}_{\boldsymbol{\Lambda}}\right)$ is already defined for the current SSLB model in \eqref{eqn:prob_lambda_given_tau}, \eqref{eqn:prob_tau_given_gamma_tilde}, and \eqref{eqn:prob_gamma_tilde_given_theta_tilde}.

For $p (\mathbf{Y} \mid \boldsymbol{\Theta}_{\mathbf{Y}} ) $ we will adopt a multinomial logistic regression approach which will allow us to define,
\begin{equation}
\label{eqn:mult_log_reg_prob}
    p \left(\mathbf{Y} \mid \boldsymbol{\Theta}_{\mathbf{Y}} \right) \propto \prod_{i=1}^N \prod_{l=1}^C \left[\exp \left(\mathbf{w}_l^T \tilde{\boldsymbol{\gamma}}_i^{\prime}\right) / \sum_{l^{\prime}=1}^c \exp \left(\mathbf{w}_{l^{\prime}}^T \tilde{\boldsymbol{\gamma}}_i^{\prime}\right)\right]^{y_{i l}} + \exp \left( - \frac{1}{2} \lambda_{\text{W}} \lVert \mathbf{W} \rVert_{F}^2 \right),
\end{equation}
where
\begin{itemize}
    \item $\boldsymbol{\Theta}_{\mathbf{Y}} = \{ \tilde{\boldsymbol{\Gamma}}^{\prime}, \mathbf{W}, \lambda_{\text{W}} \}$.
    \item $y_{i l}$ corresponds to the presence/absence (i.e., $1$ or $0$) of the $l$ disease in the $i$ sample.
    \item $C$ is the number of diseases (i.e., the number of categories in the multinomial logistic regression model) in the study.
    \item $\mathbf{W} \in \mathbb{R}^{(K + 1) \times C}$ is the matrix of weights of the multinomial logistic regression, which includes the bias coefficients. It is the main element that allows $\mathbf{Y}$ to assist in the task of assigning samples (and genes) to biclusters.
    \item $\tilde{\boldsymbol{\gamma}}_i^{\prime} = \left[ 1, \tilde{\gamma}_{i1}, \ldots, \tilde{\gamma}_{iK} \right]$.
    \item A reference category (e.g., class or disease $C$) needs to be chosen such that the column of the matrix $\mathbf{W}$ corresponding to the selected reference category has only zeros (e.g., $\mathbf{w}_C = \left[0, \ldots, 0\right]$ ).
    \item To avoid overfitting, we have introduced an $\ell_2$ regularisation term for the weight matrix $\mathbf{W}$ with regularisation hyperparameter $\lambda_{\text{W}} \in \mathbb{R}^+$.
\end{itemize}

By incorporating this approach into the model, the complete log posterior \eqref{eqn:log_post_no_prof_reg} and the expression associated with the E step \eqref{eqn:exp_log_post_no_prof_reg} will undergo minor modifications. Refer to \eqref{eqn:log_post_w_prof_reg} and \eqref{eqn:exp_log_post_w_prof_reg} in the Appendix for further details.

Below we will explain the computation of the new expectation and maximisation steps introduced by the profile regression model adapted to the SSLB model, particularly the computation of $\left\langle\widetilde{\boldsymbol{\gamma}}_i^{\prime}\right\rangle$, the second last term of \eqref{eqn:exp_log_post_w_prof_reg} that involves the computation of an expectation of a log-sum-exp expression, the estimation of $\lambda_{\text{W}}$ and the maximisation of $\mathbf{W}$. See \cite[Section 2.3]{10.1214/20-AOAS1385} and its supplementary material for details on the computation of the remaining parameters that are not affected by the introduction of profile regression to the SSLB model.

\subsection{Expectation of \texorpdfstring{$\tilde{\boldsymbol{\gamma}}_i$}{gamma tilde i}}
\label{subsec:expectation_gamma_tilde_i}

For the expectation of the $\widetilde{\gamma}_{i k}$ variables, we have

\begin{equation*}
\left\langle\widetilde{\gamma}_{ik}\right\rangle =p\left(\widetilde{\gamma}_{ik}=1 \mid \mathbf{Y}, \mathbf{T}, \widetilde{\boldsymbol{\theta}}, \mathbf{W}\right) =\frac{1}{1 + \frac{p\left(y_{ic} \mid \widetilde{\gamma}_{ik}=0, \mathbf{w}_c\right) p\left(\tau_{ik} \mid \widetilde{\gamma}_{ik}=0\right) p\left(\widetilde{\gamma}_{ik}=0 \mid \widetilde{\theta}_k\right)}{p\left(y_{ic} \mid \widetilde{\gamma}_{ik}=1, \mathbf{w}_c\right) p\left(\tau_{ik} \mid \widetilde{\gamma}_{ik}=1\right) p\left(\widetilde{\gamma}_{ik}=1 \mid \widetilde{\theta}_k\right)}}
\end{equation*}
where $c$ corresponds to the $c$-th disease presented in sample $i$, and $\mathbf{w}^c$ is the $c$-th column of the matrix of weights $\mathbf{W} \in \mathbb{R}^{(K + 1) \times C}$. From this, the only new expression left to compute is $P\left(y_{ic} \mid \widetilde{\gamma}_{ik}=1, \mathbf{w}^c\right)$. Since there is no closed-form solution, this probability will be estimated in the following way

\begin{itemize}
    \item Generate $M$ samples from $p\left(\widetilde{\gamma}_{ik}=1 \mid \mathbf{T}, \widetilde{\boldsymbol{\theta}}\right)$, a probability to which we have access (see the supplementary material of \cite{10.1214/20-AOAS1385}). This is done for every $k \in \{1, \ldots, K\}$. The results will be stored in a matrix $\mathbf{V} \in \mathbb{R}^{M\times K}$.
    \item For each column $k \in \{1, \ldots, K\}$ in $\mathbf{V}$:
    \begin{enumerate}
        \item Extract only the rows of $\mathbf{V}$ whose $k$-th column is equal to $1$, This subset of $\mathbf{V}$ can be defined as $\mathbf{V}^{\prime} \in \mathbb{R}^{M^{\prime} \times K}$ where $M^{\prime} \leq M$.
        \item Compute $p\left(y_{ic} \mid \mathbf{V}^{\prime}, \mathbf{w}^c \right)$ using \eqref{eqn:mult_log_reg_prob}. The result of this computation can be stored in a vector $\mathbf{q} \in \mathbb{R}^{M^{\prime}}$.
        \item Finally, estimate $p\left(y_{ic} \mid \widetilde{\gamma}_{ik}=1, \mathbf{w}^c\right)$ as follows
        \begin{equation*}
            p\left(y_{ic} \mid \widetilde{\gamma}_{ik}=1, \mathbf{w}^c\right) \approx \frac{\sum_{i = 1}^{M^{\prime}} q_i}{M}.
        \end{equation*}
    \end{enumerate}
\end{itemize}

In our numerical experiments, we have empirically observed that using $M=50$ results in estimates of $\left\langle\widetilde{\gamma}_{ik}\right\rangle$ with sufficiently low variance and consistent results across multiple trials, indicating that this choice of $M$ is sufficient for reliable estimation.

\subsection{Expectation of the log-sum-exp expression in \texorpdfstring{$Q(\boldsymbol{\Delta} \mid \boldsymbol{\Delta}^{(t)})$}{Q(∆|∆t)}}
\label{subsec:expect_log_sum_exp}

For the computation of the last term of \eqref{eqn:exp_log_post_w_prof_reg} which implies an expectation of a log-sum-exp expression, since we now have a way to estimate $\left\langle\widetilde{\gamma}_{ik}\right\rangle$, the computation of this expectation is a simple Monte Carlo estimate as follows:

\begin{itemize}
    \item Generate \( \widetilde{\boldsymbol{\gamma}}_i^{\prime (1)}, \ldots, \widetilde{\boldsymbol{\gamma}}_i^{\prime (m)} \) samples from $p\left(\widetilde{\gamma}_{ik}=1 \mid \mathbf{Y}, \mathbf{T}, \widetilde{\boldsymbol{\theta}}, \mathbf{W}\right)$ previously estimated in Section \ref{subsec:expectation_gamma_tilde_i}, for each $k \in \{1, \ldots, K \}$.
    \item Compute
    \begin{equation*}
        \Biggl\langle \log \left[\sum_{l=1}^C \exp \left(\mathbf{w}_l^T \widetilde{\boldsymbol{\gamma}}_i^{\prime}\right)\right] \Biggr\rangle \approx \frac{1}{m} \sum_{i=1}^{m} \log \left[ \sum_{l=1}^C \exp \left( \mathbf{w}_l^T \widetilde{\boldsymbol{\gamma}}_i^{\prime (m)}  \right) \right]
    \end{equation*}
\end{itemize}

\subsection{Estimation of hyperparameter \texorpdfstring{$\lambda_{\text{W}}$}{λW} }
To estimate the regularisation hyperparameter $\lambda_{\text{W}}$ of the $\ell_2$ penalisation term of the multinomial logistic regression weights matrix $\mathbf{W}$, we adopted an empirical Bayesian approach by maximum marginal likelihood estimate. This can be done by solving the following

\begin{align}
    \lambda_{\text{W}}^{\mbox{*}} & = \argmax_{\lambda_{\text{W}}} p \left(\mathbf{Y} \mid \tilde{\boldsymbol{\Gamma}}^{\prime}, \lambda_{\text{W}} \right) \nonumber \\ &  = \argmax_{\lambda_{\text{W}}} \int_{\mathbb{R}^{(K + 1) \times C}} \prod_{i=1}^N \prod_{l=1}^C\left[\exp \left(\mathbf{w}_l^T \tilde{\boldsymbol{\gamma}}_i^{\prime}\right) / \sum_{l^{\prime}=1}^c \exp \left(\mathbf{w}_{l^{\prime}}^T \tilde{\boldsymbol{\gamma}}_i^{\prime}\right)\right]^{y_{i l}} + \exp \left( - \frac{1}{2} \lambda_{\text{W}} \lVert \mathbf{W} \rVert_{F}^2 \right) d \mathbf{W} .  \label{eqn:mmle_integral}
\end{align}

Since the latter integral, i.e., the resulting marginal likelihood of the multinomial logistic regression model, is computationally intractable, we will apply the Stochastic Optimisation via Unadjusted Langevin (SOUL) method \cite{de2021efficient}, which is specifically designed for this type of problem. We will explain this method in detail below.

We can solve \eqref{eqn:mmle_integral} iteratively using the projected gradient algorithm \cite[Section 5]{LEVITIN19661}
\begin{equation}
    \lambda_{\text{W}}^{(n+1)} = \Pi_{\Theta_{\lambda_{\text{W}}}} \left[ \lambda_{\text{W}}^{(n)} + \delta_{\text{PGA}}^{(n)} \nabla_{\lambda_{\text{W}}} p \left(\mathbf{Y} \mid \tilde{\boldsymbol{\Gamma}}^{\prime}, \lambda_{\text{W}}^{(n)} \right) \right],
    \label{eqn:projected_grad_alg}
\end{equation}
by computing a sequence $(\lambda_{\text{W}}^{(n)})_{n \in \mathbb{N}}$ associated with the latter recursion, where $\Pi_{\Theta_{\lambda_{\text{W}}}}$ denotes the projection onto the compact convex set $\Theta_{\lambda_{\text{W}}} \subset (0,+\infty)$ and $(\delta_{\text{PGA}}^{(n)})_{n \in \mathbb{N}}$ is a sequence of non-increasing step sizes\footnote{When denoting sequences of values, we use superscripts in parentheses instead of subscript and superscript notation to avoid confusion with matrix notation. For example, \( \mathbf{M}^{(i)} \) represents the \( i \)-th value in a sequence, rather than a power or column/row vector.}. However, the gradient in \eqref{eqn:projected_grad_alg} is intractable, as we saw in \eqref{eqn:mmle_integral}. For this case, we can replace this gradient with a stochastic estimator by applying Fisher's identity \cite[Section D.2]{douc2014nonlinear}

\begin{align*}
    \nabla_{\lambda_{\text{W}}} p \left(\mathbf{Y} \mid \tilde{\boldsymbol{\Gamma}}^{\prime}, \lambda_{\text{W}} \right) & = \int_{\mathbb{R}^{(K + 1) \times C}} \frac{\nabla_{\lambda_{\text{W}}} p \left(\mathbf{W}, \mathbf{Y} \mid \tilde{\boldsymbol{\Gamma}}^{\prime}, \lambda_{\text{W}} \right)}{p \left(\mathbf{W}, \mathbf{Y} \mid \tilde{\boldsymbol{\Gamma}}^{\prime}, \lambda_{\text{W}} \right)} p \left(\mathbf{W} \mid \mathbf{Y},  \tilde{\boldsymbol{\Gamma}}^{\prime}, \lambda_{\text{W}} \right) d \mathbf{W} \\ & = \int_{\mathbb{R}^{(K + 1) \times C}} \nabla_{\lambda_{\text{W}}} \log p \left(\mathbf{W}, \mathbf{Y} \mid \tilde{\boldsymbol{\Gamma}}^{\prime}, \lambda_{\text{W}} \right) p \left(\mathbf{W} \mid \mathbf{Y},  \tilde{\boldsymbol{\Gamma}}^{\prime}, \lambda_{\text{W}} \right) d \mathbf{W}, 
\end{align*}
where $p (\mathbf{W} \mid \mathbf{Y},  \tilde{\boldsymbol{\Gamma}}^{\prime}, \lambda_{\text{W}} )$ is the posterior distribution of $\mathbf{W}$, given by
\begin{equation*}
    p \left(\mathbf{W} \mid \mathbf{Y},  \tilde{\boldsymbol{\Gamma}}^{\prime}, \lambda_{\text{W}} \right) \propto p \left(\mathbf{Y} \mid \mathbf{W}, \tilde{\boldsymbol{\Gamma}}^{\prime}, \lambda_{\text{W}} \right) p \left(\mathbf{W} \mid \lambda_{\text{W}} \right).
\end{equation*}
Given the fact that $p (\mathbf{W}, \mathbf{Y} \mid \tilde{\boldsymbol{\Gamma}}^{\prime}, \lambda_{\text{W}} ) = p (\mathbf{Y} \mid \mathbf{W}, \tilde{\boldsymbol{\Gamma}}^{\prime}) p (\mathbf{W} \mid \lambda_{\text{W}})$ we have

\begin{equation*}
    \nabla_{\lambda_{\text{W}}} p \left(\mathbf{Y} \mid \tilde{\boldsymbol{\Gamma}}^{\prime}, \lambda_{\text{W}} \right) = \int_{\mathbb{R}^{(K + 1) \times C}} \nabla_{\lambda_{\text{W}}} \log p \left(\mathbf{W} \mid \lambda_{\text{W}} \right) p \left(\mathbf{W} \mid \mathbf{Y},  \tilde{\boldsymbol{\Gamma}}^{\prime}, \lambda_{\text{W}} \right) d \mathbf{W}
\end{equation*}
where
\begin{align*}
    \log p \left(\mathbf{W} \mid \lambda_{\text{W}} \right) & = - \frac{1}{2} \lambda_{\text{W}} \lVert \mathbf{W} \rVert_{F}^2 - \int_{\mathbb{R}^{(K + 1) \times C}} \exp \left( - \frac{1}{2} \lambda_{\text{W}} \lVert \mathbf{W} \rVert_{F}^2 \right) d \mathbf{W} \\ & = - \frac{1}{2} \lambda_{\text{W}} \lVert \mathbf{W} \rVert_{F}^2 - \log \left( \frac{2 \pi}{\lambda_{\text{W}}} \right)^{0.5 \times (K + 1) \times C}.
\end{align*}
Therefore
\begin{equation*}
    \nabla_{\lambda_{\text{W}}} \log p \left(\mathbf{W} \mid \lambda_{\text{W}} \right) = -0.5 \lVert \mathbf{W} \rVert_{F}^2 + \frac{(K + 1) \times C}{2 \lambda_{\text{W}}}
\end{equation*}
Having finally that
\begin{align*}
    \nabla_{\lambda_{\text{W}}} p \left(\mathbf{Y} \mid \tilde{\boldsymbol{\Gamma}}^{\prime}, \lambda_{\text{W}} \right) & = \frac{(K + 1) \times C}{2 \lambda_{\text{W}}} - 0.5 \int_{\mathbb{R}^{(K + 1) \times C}} \lVert \mathbf{W} \rVert_{F}^2 p \left(\mathbf{W} \mid \mathbf{Y},  \tilde{\boldsymbol{\Gamma}}^{\prime}, \lambda_{\text{W}} \right) d \mathbf{W} \\ & =  \frac{(K + 1) \times C}{2 \lambda_{\text{W}}} - \frac{1}{2} \mathbb{E}_{\mathbf{W} \mid \mathbf{Y},  \tilde{\boldsymbol{\Gamma}}^{\prime}, \lambda_{\text{W}}} \left[ \lVert \mathbf{W} \rVert_{F}^2 \right],
\end{align*}
that is, the gradient we need for the iterative scheme in \eqref{eqn:projected_grad_alg} depends on the computation of an expectation that can be estimated using MCMC methods. To approximate samples from the posterior distribution $p (\mathbf{W} \mid \mathbf{Y}, \tilde{\boldsymbol{\Gamma}}^{\prime}, \lambda_{\text{W}} )$ and compute the latter expectation, the SOUL method uses the unadjusted Langevin algorithm (ULA) \cite{bj/1178291835, 7e6dbf96-e106-36fd-b898-4be31ae7ec6e, 10.1214/16-AAP1238}, given by
\begin{equation}\label{eqn:ULA_SOUL}
    W^{(k+1)} = W^{(k)} - \delta_{\text{ULA}} \nabla_W \log  p (W^{(k)} \mid \mathbf{Y},  \tilde{\boldsymbol{\Gamma}}^{\prime}, \lambda_{\text{W}} ) + \sqrt{2 \delta_{\text{ULA}}} Z^{(n+1)} ,
\end{equation}
where $\delta_{\text{ULA}} >0$ is a given step size and $(Z^{(n+1)})_{n \geq 0}$ is an i.i.d. sequence of $(K + 1) \times C$ - dimensional standard Gaussian random vectors. We are ready to present the SOUL method adapted for this problem, which can be found in Algorithm \ref{alg:SOUL}.
\begin{algorithm}
\caption{SOUL algorithm for the estimation of $\lambda_{\text{W}}$}
\label{alg:SOUL}
\begin{algorithmic}[1]
\REQUIRE $\lambda_{\text{W}}^{(0)} \in \Pi_{\Theta_{\lambda_{\text{W}}}}$, $W^{(0,0)} \in \mathbb{R}^{(K + 1) \times C}$, $\delta_{\text{ULA}}, \delta_{\text{PGA}}^{(0)} \in \mathbb{R}$, $m, n \in \mathbb{N}$
\FOR{$i = 1$ \textbf{to} $n$}
    \IF{$i > 1$}
        \STATE{Set $W^{(i,0)} = W^{(i-1, m)}$},
    \ENDIF
    \FOR{$j = 0$ \textbf{to} $m - 1$}
        \STATE $Z^{(i,j+1)} \sim N(0, \mathbf{I}_{(K + 1) \times C})$
        \STATE $W^{(i,j+1)} = W^{(i,j)} - \delta_{\text{ULA}} \nabla_W \log  p (W^{(i,j)} \mid \mathbf{Y},  \tilde{\boldsymbol{\Gamma}}^{\prime}, \lambda_{\text{W}}^{(i-1)} ) + \sqrt{2 \delta_{\text{ULA}}} Z^{(i,j+1)}$
    \ENDFOR
    \STATE $\lambda_{\text{W}}^{(i)} = \Pi_{\Theta_{\lambda_{\text{W}}}} \left[ \lambda_{\text{W}}^{(i-1)} + \frac{\delta_{\text{PGA}}^{(i-1)}}{2m} \sum_{j=1}^m \left\lbrace  \frac{(K + 1) \times C}{\lambda_{\text{W}}^{(i-1)}} - \lVert W^{(i,j)} \rVert_{F}^2 \right\rbrace \right]$
\ENDFOR
\ENSURE $\hat{\lambda}_{\text{W}}^{(n)} = \sum_{i=1}^{n} w^{(i)} \lambda_{\text{W}}^{(i)} / \sum_{i=1}^{n} w^{(i)}$
\end{algorithmic}
\end{algorithm}
\subsubsection{SOUL implementation guidelines}
The implementation guidelines and details about the SOUL algorithm can be found in \cite{de2021efficient} and in \cite[Section 3.3]{doi:10.1137/20M1339829}. For completeness, we will provide some details below.

\paragraph{Setting \texorpdfstring{$\delta_{\text{PGA}}^i$}{δPGA}, and \texorpdfstring{$m$}{m}}
It is suggested in \cite{doi:10.1137/20M1339829} to set $\delta_{\text{PGA}}^{(i)} = C_0 i^{-p}$ where $p$ is within the range $[0.6, 0.9]$ (in our experiments, we set $p=0.8$) and $C_0 \in \mathbb{R}$ a constant that can be initially set as $(\lambda_{\text{W}}^{(0)} \times (K + 1) \times C)^{-1}$ and adjusted as needed. For $m$, we followed the recommendation in \cite{de2021efficient, doi:10.1137/20M1339829} using a single sample per iteration (that is, $m=1$), as we did not observe significant differences with larger values of $m$.

\paragraph{Setting \texorpdfstring{$w^{(i)}$}{wi}}

According to the guidelines in \cite{doi:10.1137/20M1339829}, it is recommended to set $w^{(i)}$ as follows:
\begin{equation*}
w^{(i)} =
\begin{cases}
0 & \text{if } i < N_0, \\
1 & \text{if } N_0 \leq i \leq n, \\
\end{cases}
\end{equation*}
where $N_0 \in \mathbb{N}$ is the number of initial iterations to be discarded to reduce non-asymptotic bias, which corresponds to a burn-in stage. The range $i \in [N_0, n]$ represents the estimation phase of the averaging where the values of $\lambda_{\text{W}}^{(i)}$ have reached convergence and stabilised. In the interest of computational efficiency, we have set in our experiments $N_0 = 75$ and $n = 150$.

\paragraph{Implementation in Logarithmic Scale}
The proposed methods for estimating $\lambda_{\text{W}}$ generally achieve better numerical convergence when implemented on a logarithmic scale, as recommended in \cite[Section 3.3.2]{doi:10.1137/20M1339829}. Therefore, we apply a variable transformation $\kappa = \log(\lambda_{\text{W}})$, estimate $\hat{\kappa}$ using the SOUL algorithm, and then determine $\hat{\lambda}_{\text{W}} = e^{\hat{\kappa}}$.

This variable transformation necessitates a slight modification in the gradient calculations, which must be scaled by $e^{\kappa^{(n)}}$ to adhere to the chain rule. For instance, step 9 in Algorithm \ref{alg:SOUL} is updated to \[\kappa^{(i+1)} = \Pi_{\Theta_{\kappa}} \left[ \kappa^{(i)} + e^{\kappa^{(i)}} \frac{\delta_{\text{PGA}}^{(i+1)}}{2m} \sum_{j=1}^{m} \left\lbrace \left( (K + 1) \times C \times e^{-\kappa^{(i)}} \right) - \lVert W^{(i,j)} \rVert_{F}^2 \right\rbrace \right],\] where $\Theta_{\kappa} = \{\log(\lambda_{\text{W}}) : \lambda_{\text{W}} \in \Theta_{\lambda_{\text{W}}}\}$ represents the permissible range of $\kappa$ values taken logarithmically.

\subsection{Gradient of ridge multinomial logistic regression}
\label{subsec:gradient_mult_log_rigde_reg}
In the ULA step of the SOUL method and the computation of $\mathbf{\hat{W}}$ which will be further explained, we need to compute the gradient of the logarithm of the ridge multinomial logistic regression model, that is
\begin{equation*}
    \nabla_{\mathbf{w}} \log p (\mathbf{Y} \mid \tilde{\boldsymbol{\Gamma}}^{\prime}, \mathbf{W}, \hat{\lambda}_{\text{W}} ) = \nabla_{\mathbf{w}} \left[ \sum_{i=1}^N \sum_{l=1}^C y_{i l} \mathbf{w}_l^T\left\langle\widetilde{\boldsymbol{\gamma}}_i^{\prime}\right\rangle + \sum_{i=1}^N \Biggl\langle \log \left[\sum_{l=1}^C \exp \left(\mathbf{w}_l^T \widetilde{\boldsymbol{\gamma}}_i^{\prime}\right)\right] \Biggr\rangle  + \frac{1}{2} \hat{\lambda}_{\text{W}} \lVert \mathbf{W} \rVert_{F}^2 \right].
\end{equation*}
This gradient is a well-known result in the literature (see, e.g., \cite[Ex. 4.4]{Hastie2009}). Since it involves an expectation that does not have a closed-form solution, we must compute a Monte Carlo estimate of this log-sum-exp expectation term beforehand. This process was previously explained in Section \ref{subsec:expect_log_sum_exp}. The gradient is then given by
\begin{align*}
    \nabla_{\mathbf{w}} \log p (\mathbf{Y} \mid \tilde{\boldsymbol{\Gamma}}^{\prime}, \mathbf{W}, \hat{\lambda}_{\text{W}} ) & \approx - \frac{1}{J} \sum_{j=1}^J \widetilde{\boldsymbol{\Gamma}}^{\prime (j) \mathsf{T}} \left[ p \left(\mathbf{Y} \mid \tilde{\boldsymbol{\Gamma}}^{\prime (j)}, \mathbf{W}, \hat{\lambda}_{\text{W}} \right) - \mathbf{Y}\right] + \hat{\lambda}_{\text{W}} \mathbf{W} \\ & \coloneq \overline{\nabla}_{\mathbf{w}} \log p (\mathbf{Y} \mid \tilde{\boldsymbol{\Gamma}}^{\prime}, \mathbf{W}, \hat{\lambda}_{\text{W}} ),
\end{align*}
where we have generated a collection of $\tilde{\boldsymbol{\Gamma}}^{\prime (1)}, \ldots, \tilde{\boldsymbol{\Gamma}}^{\prime (J)}$ samples from $p ( \widetilde{\gamma}_{ik}=1 \mid \mathbf{Y}, \mathbf{T}, \widetilde{\boldsymbol{\theta}}, \mathbf{W} )$ to compute the Monte Carlo estimate, as explained in Section \ref{subsec:expect_log_sum_exp}. In addition, $p (\mathbf{Y} \mid \tilde{\boldsymbol{\Gamma}}^{\prime}, \mathbf{W}, \hat{\lambda}_{\text{W}} )$ is defined in \eqref{eqn:mult_log_reg_prob}. We have empirically found that $T = 30$ is enough to reach a good accuracy level while maintaining a low computational cost.

\subsection{Maximisation step regarding the variable \texorpdfstring{$\mathbf{W}$}{W} }
Finally, the last expression to compute in the new SSLB model is given by
\begin{equation*}
    \widehat{\mathbf{W}}=\underset{\mathbf{W} \in \mathbb{R}^{(K+1) \times C}}{\operatorname{argmin}} \sum_{i=1}^N \sum_{l=1}^C y_{i l} \mathbf{w}_l^T\left\langle\widetilde{\boldsymbol{\gamma}}_i^{\prime}\right\rangle+\sum_{i=1}^N \Biggl\langle \log \left[\sum_{l=1}^C \exp \left(\mathbf{w}_l^T \widetilde{\boldsymbol{\gamma}}_i^{\prime}\right)\right] \Biggr\rangle + \frac{1}{2} \hat{\lambda}_{\text{W}} \lVert \mathbf{W} \rVert_{F}^2.
\end{equation*}
Since there is no close-form solution for the latter, we decided to implement an accelerated gradient descent (AGD) algorithm \cite{1370862715914709505, doi:10.1137/0802032, salzo2012inexact} to ensure rapid convergence to the minimum. To apply AGD, we need the gradient of the latter expression, which was given in Section \ref{subsec:gradient_mult_log_rigde_reg}. This allows us to define the following iterative scheme
\[
\mathbf{W}^{(0)} = \mathbf{W}^{(-1)} = \mathbf{0} \in \mathbb{R}^{(K+1)\times C}; \; \mathbf{V}^{(0)} \in \mathbb{R}^{(K+1)\times C}; \; t_0 = 0 \in \mathbb{R},
\]
\begin{align*}
    t_{s+1} & =\frac{1+\sqrt{1+4 t_s^2}}{2}, \\
    \mathbf{V}^{(s)} & =\mathbf{W}^{(s)}+\frac{t_s-1}{t_{s+1}}\left(\mathbf{W}^{(s)}-\mathbf{W}^{(s-1)}\right), \\
    \mathbf{W}^{(s+1)} & =\mathbf{V}^{(s)} + \delta_{\text{AGD}} \overline{\nabla}_{\mathbf{w}} \log p (\mathbf{Y} \mid \tilde{\boldsymbol{\Gamma}}^{\prime}, \mathbf{V}^{(s)}, \hat{\lambda}_{\text{W}} ),
\end{align*}
\[
s \in \{0,\ldots, S - 1\} \subset \mathbb{N},
\]
where $\delta_{\text{AGD}}$ is the step size or learning rate of the iterative AGD scheme, which must be carefully set to avoid divergence. Note that, on each AGD iteration, we need to generate a collection of $\tilde{\boldsymbol{\Gamma}}^{\prime (1)}, \ldots, \tilde{\boldsymbol{\Gamma}}^{\prime (J)}$ samples from $p ( \widetilde{\gamma}_{ik}=1 \mid \mathbf{Y}, \mathbf{T}, \widetilde{\boldsymbol{\theta}}, \mathbf{W} )$ to compute the Monte Carlo estimate of the gradient. 

\subsection{Step size \texorpdfstring{$\delta_{\text{ULA}}$}{δ ULA}  \& \texorpdfstring{$\delta_{\text{AGD}}$}{δ AGD}}
Determining suitable step sizes for the ULA and AGD algorithms is essential for guaranteeing convergence and computational efficiency. The literature offers well-established guidance on selecting the step size \(\delta_{\text{AGD}}\). Specifically, \(\delta_{\text{AGD}} \leq 1 / L_f \) where \(L_f\) is the Lipschitz constant of $\nabla_{\mathbf{w}} \log p (\mathbf{Y} \mid \tilde{\boldsymbol{\Gamma}}^{\prime}, \mathbf{W}, \hat{\lambda}_{\text{W}} )$. 

The value of \(L_f\) can be obtained from the Hessian of the objective function. For multinomial logistic regression \cite{bohning1992multinomial}, \(L_f\) is given by
\begin{equation*}
    L_f  = \lambda_{\max} \left[ \frac{1}{2} \left( \textbf{I}_C - \mathbf{1}_C \mathbf{1}_C^T / C \right) \otimes \tilde{\boldsymbol{\Gamma}}^{\prime T} \tilde{\boldsymbol{\Gamma}}^{\prime} \right] + \lambda_{\text{W}},
\end{equation*}
where $C$ is the number of classes of the multinomial logistic regression model. This ensures that the AGD update step remains within a stable region, facilitating steady progress towards the optimal solution.

For the ULA in the SOUL method, when $\nabla_{\mathbf{w}} \log p (\mathbf{Y} \mid \tilde{\boldsymbol{\Gamma}}^{\prime}, \mathbf{W}, \hat{\lambda}_{\text{W}} )$ is Lipschitz continuous with Lipschitz constant $L_f$, it has been shown that \(\delta_{\text{ULA}} \in (0, 2 / L_f] \) ensures that the Markov chain $(W^{(k)})_{k \geq 0}$ described in \eqref{eqn:ULA_SOUL} is ergodic with stationary distribution close to the true target distribution \cite{10.1214/16-AAP1238}. Therefore, following both AGD and ULA specifications, we have decided to set \(\delta_{\text{AGD}} = \delta_{\text{ULA}} = 0.95 / L_f\).

With the methodology established, we will refer to this modified version of the SSLB model as \textbf{OG-SSLB}, which stands for \textbf{Outcome-Guided Spike-and-Slab Lasso Biclustering}.

\section{Numerical Results}
\label{sec:numerical_results}

\subsection{Simulation Study}
In this section, we evaluate the performance of OG-SSLB compared to SSLB in a simulation setting. We use the consensus score metric \cite{10.1093/bioinformatics/btq227} to measure the accuracy of biclusters identified by each method relative to the true biclusters. The highest possible consensus score is 1, indicating identical sets of biclusters.

We reproduce the simulation described in \cite[Section 3.1]{10.1214/20-AOAS1385}, where a simulated dataset with $N = 300$, $G = 1000$, and $K = 15$ biclusters is examined. The data simulation follows settings closely aligned with those in the FABIA \cite{10.1093/bioinformatics/btq227} and SSLB studies. The data matrix $\mathbf{X}$ is generated as $\mathbf{\Lambda Z}^T + \mathbf{E}$, with each entry in the noise matrix $\mathbf{E}$ sampled from an independent standard normal distribution. For each column $\boldsymbol{\lambda}^k$, the number of samples in bicluster $k$ is drawn uniformly from $\{5, \ldots, 20\}$. The indices of these elements are randomly selected and assigned values from $N(\pm 2, 1)$, with the sign of the mean chosen randomly. The elements of $\boldsymbol{\lambda}^k$ not in the biclusters have values drawn from $N(0, 0.2^2)$. The columns $\mathbf{z}^k$ are generated similarly, except that the number of elements in each bicluster is drawn from $\{10, \ldots, 50\}$.

To construct the disease outcome matrix $\mathbf{Y}$ for the OG-SSLB algorithm, we first generate a matrix of weights $\mathbf{W}$ in the following way
\begin{enumerate}
    \item \textbf{Intercepts (Baseline Weights for Healthy Control Group)}: The first row of \( \mathbf{W} \), representing the intercepts, is initialised with small values
    \[
    W_{1j} = \log (\epsilon) \quad \text{for all } j,
    \]
    where \(0 < \epsilon < 1 \). These small values correspond to the reference class (i.e., the healthy control group, HC) in the multinomial logistic regression model. In logistic regression, the intercept term controls the baseline probability of a sample belonging to the reference class when no covariates (in this case, the bicluster assignments) are active. By assigning a small value to \( W_{1j} \), we increase the baseline probability for healthy control samples when no bicluster is assigned to a sample. Since \( \log(\epsilon) \) with \( \epsilon < 1 \) results in a negative value, this translates into a higher probability of belonging to the reference class (HC).
    \item \textbf{Weight Assignment for Biclusters (Bias Towards Disease Samples)}: For samples belonging to a bicluster, we adjust the weights in the remaining rows of \( \mathbf{W} \), corresponding to the non-reference classes (i.e., the disease classes). Specifically, we set the weights for the non-reference class as:
    \[
    W_{ij} = \log(1 / \epsilon) \quad \text{for } i>1, j .
    \]
    Here, \( \log(1/\epsilon) \), where \( \epsilon < 1 \), results in a positive value, increasing the likelihood that samples assigned to a bicluster are classified as disease samples. Importantly, only one column \( j \) is randomly selected for each row \( i \) in \( \mathbf{W} \) to be assigned this value. This ensures that a specific bicluster is more strongly associated with a particular disease group, effectively biasing the model toward classifying samples in that bicluster as disease samples.
\end{enumerate}
Finally, the assignment of disease for each sample is determined by applying the multinomial logistic regression model using the defined weights \( \mathbf{W} \).

The rationale behind making samples belonging to a bicluster more likely to be classified as disease cases stems from the biological assumption that disease states are often driven by specific gene expression patterns \cite{ota2021dynamic, MESKO2011223, doi:10.1056/NEJMoa021967}. Biclustering aims to identify subsets of genes that co-vary together in certain subsets of samples, which may represent distinct biological processes or pathways that are activated in disease conditions.

We adopt similar hyperparameter configurations for SSLB and OG-SSLB, detailed in \cite[Section 2.5]{10.1214/20-AOAS1385}. In particular, the slab parameters for the loadings and the factors, $\mathbf{Z}$ and $\boldsymbol{\Lambda}$, are set to $\omega_1, \tilde{\omega}_1 = 1$. The spike parameters for $\mathbf{Z}$ follow an increasing sequence of $\omega_0 \in \{ 1, 5, 10, 50, 100, 500, 10^3, 10^4, 10^5, 10^6, 10^7 \}$. The spike parameters for $\boldsymbol{\Lambda}$ are chosen as $\tilde{\omega}_0 \in \{ 1, 5, \ldots, 5 \}$ to correspond to the length of the sequence $\omega_0$. Specifically, the values of $\tilde{\omega}_0$ are fixed at $\tilde{\omega}_0 = 5$. Furthermore, the initial overestimate of the number of biclusters is set to $K^* = 30$.

We compared 50 realisations of SSLB and OG-SSLB using the same simulated dataset across all runs while varying the algorithmic initial conditions for each of the fifty runs. This analysis was performed under three distinct implementations: SSLB/OG-SSLB with the Pitman–Yor extension (PY), where $\tilde{\alpha} = 1$ and $d = 0.5$, SSLB/OG-SSLB with the stick-breaking IBP prior (IBP) where $\tilde{\alpha} = 1$, and SSLB/OG-SSLB with the finite approximation to the IBP prior (Beta-Binomial, BB), where $\tilde{a} = 1/K^*$ and $\tilde{b} = 1$. For OG-SSLB, we implemented two variations: a non-informative approach, which assigns values around $\log(1)$ to all elements of the matrix $\mathbf{W}$, resulting in an imprecise simulated $\mathbf{Y}$ outcome, and an informative approach, which assigns $\log(1/4)$ to the intercepts and $\log(4)$ to specific bicluster-disease elements (i.e., one specific column in each row of $\mathbf{W}$), yielding a more informative simulated $\mathbf{Y}$ outcome. We aim to show the difference in adding more information to the model.

\begin{figure}[ht]
\centering
\includegraphics[width = 0.75\linewidth]{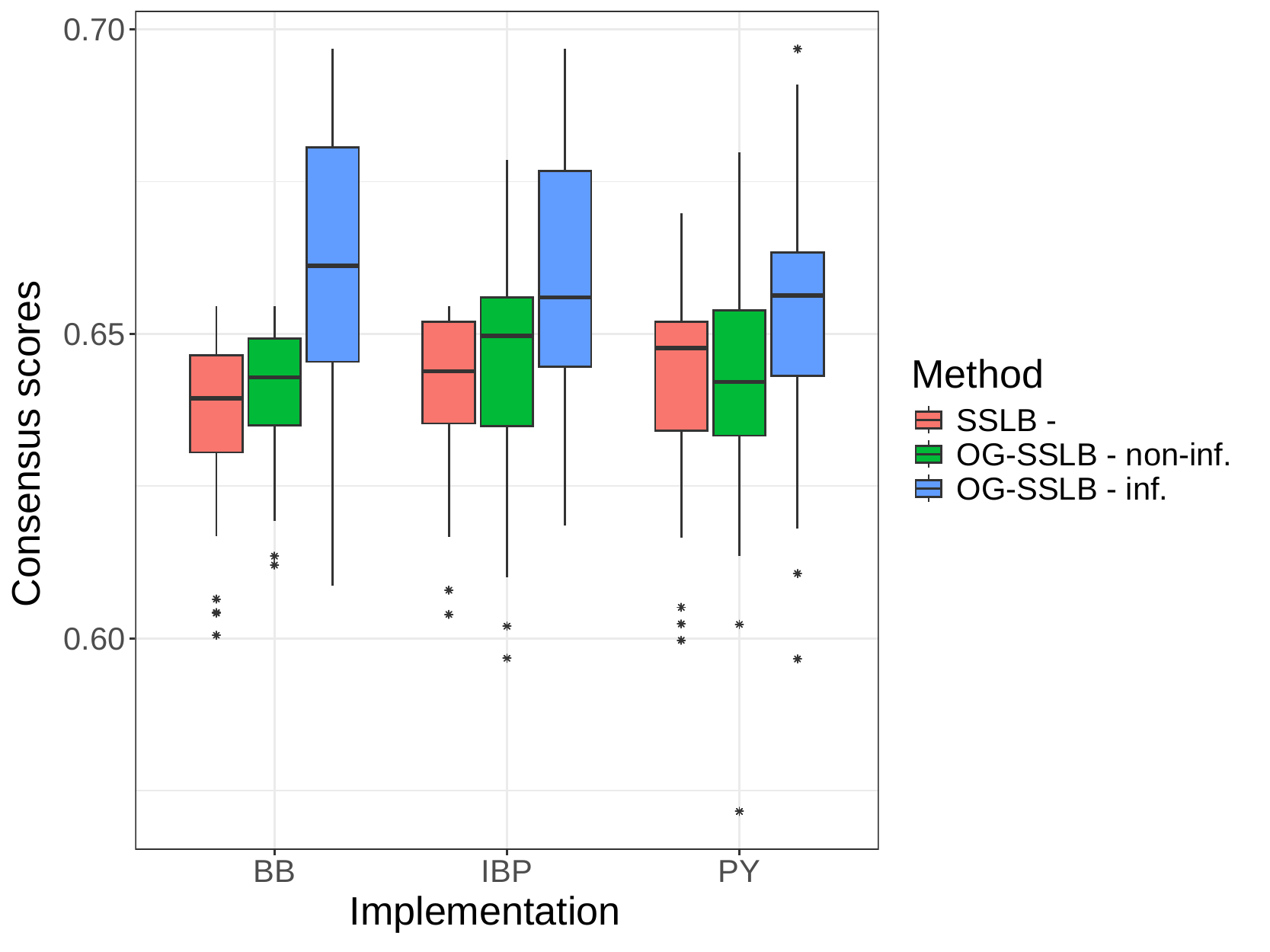}
\caption{Consensus scores of 50 replications of SSLB and OG-SSLB, both using three different prior implementations for $\tilde{\boldsymbol{\Gamma}}$: BB, PY and IBP (see Section \ref{subsubsec:sslb_priors} for details). For OG-SSLB, we implement a non-informative (i.e., non-inf.) approach, setting values for all elements of the matrix $\mathbf{W}$ around $\log(1)$, which produces a diffusive $\mathbf{Y}$ outcome, and an informative (i.e., inf.) approach, where we set $\log(1/4)$ for the intercepts and $\log(4)$ for specific bicluster-disease elements in the matrix $\mathbf{W}$, resulting in a more informative $\mathbf{Y}$ outcome.}\label{fig:num_exp_normal_consensus}
\end{figure}

The distribution of the consensus score for each method can be seen in Figure \ref{fig:num_exp_normal_consensus}. OG-SSLB consistently achieves higher consensus scores in all three prior versions of the binary indicators for the factors $\boldsymbol{\Lambda}$, reaching even higher precision in the informative case (with a slight increase in the PY prior version for the informative case). Table \ref{tab:num_exp_normal_consensus} presents the estimated number of biclusters $\hat{K}$ for the methods that estimate $K$. All implementations of the informative OG-SSLB approach are closer to the true number of biclusters compared to SSLB. Regarding the corresponding run times, the SSLB implementations required between 35 and 60 seconds, whereas the OG-SSLB implementations took between 2600 and 3500 seconds.

\begin{table}[ht]
\centering
\caption{Mean estimated number of biclusters, $\hat{K}$, over 50 replications ($K = 15$).}\label{tab:num_exp_normal_consensus}
  \begin{tabular}{l|ccc}
    \toprule
    \textbf{SSLB} & \multicolumn{3}{c}{\textbf{Method}} \\ 
    \textbf{implementation} & \textbf{SSLB} & \textbf{OG-SSLB (non-inf.)} & \textbf{OG-SSLB (inf.)} \\ \midrule
    BB & 14.04 & 14.02 & 14.50 \\
    IBP & 14.10 & 14.36 & 14.50 \\
    PY & 14.34 & 14.42 & 14.46 \\
    \bottomrule
  \end{tabular}
\end{table}

\subsection{Immune Cell Gene Expression Atlas, University of Tokyo}
Detecting biclusters in transcriptomic data is one of the motivating applications for OG-SSLB, therefore, we also applied the SSLB and OG-SSLB methods to gene expression data from \cite{ota2021dynamic}. This study provides a comprehensive database of transcriptomic and genome sequencing data from a wide range of immune cells from patients with immune-mediated diseases (IMD). This collection of data, termed the ``Immune Cell Gene Expression Atlas from the University of Tokyo (ImmuNexUT)", includes gene expression patterns consisting of healthy volunteers and patients diagnosed with systemic lupus erythematosus (SLE), idiopathic inflammatory myopathy (IIM), systemic sclerosis (SSc), mixed connective tissue disease (MCTD), Sjögren’s syndrome (SjS), rheumatoid arthritis (RA), Behçet's disease (BD), adult-onset Still's disease (AOSD), ANCA-associated vasculitis (AAV), or Takayasu arteritis (TAK). The dataset encompasses 28 distinct immune cell types, nearly covering all peripheral immune cells. We anticipate that a subset of genes may cluster in a subset of patients who share some specific aetiology, and that this bicluster will be enriched in genes related to that aetiology and patients with related diseases.

In order to evaluate our method in a real-world dataset where we have some expectation of what to find, we focused on monocytes, which are known to express an interferon-response gene expression signature, found more often in patients with IMD, and particularly SLE \cite{nikolakis2023restoration, doi:10.1126/science.abf1970}. The data were pre-processed as follows:
\begin{enumerate}
    \item We perform batch normalisation using the ComBat-seq R package \cite{10.1093/nargab/lqaa078}.
    \item We reduce low-count genes using the edgeR package \cite{10.1093/bioinformatics/btp616}.
    \item We calculated the Pearson correlation matrix between genes and setting a threshold at the 90th percentile, we focus on the most highly correlated gene pairs. Genes with fewer than five other genes correlating above this threshold are removed to eliminate those with weak or non-specific interactions, which could be noisy or less informative. This step ensures that only genes with strong co-expression relationships, potentially reflecting meaningful biological connections, are retained.
    \item Finally, to correct for technical variation and differences in sequencing depth between samples, we applied the median of ratios normalisation method, as implemented in the DESeq2 R package \cite{love2014moderated}. This normalisation ensures that gene expression differences reflect true biological variability rather than artefacts from varying read counts across samples.
\end{enumerate}

Following preprocessing, we obtained a dataset comprising $N = 410$ and $G = 11215$. We ran 20 different replicates of both SSLB and OG-SSLB using the IBP prior for $\tilde{\boldsymbol{\Gamma}}$, with hyperparameters $\tilde{\alpha} = 1/N$ and $d=0$, and the Beta-Bernoulli prior hyperparameters for $\boldsymbol{\Gamma}$ set to $a = 1 / (G K^*)$ and $b = 1$. Our choice of using the IBP prior and the specified hyperparameters values for the factor and loading binary indicator matrices is in agreement with the real data experiments performed in the SSLB paper (see \cite[Sections 4 and 5]{10.1214/20-AOAS1385} for details). The remaining hyperparameter settings were similarly aligned with those of the previous numerical experiment. Furthermore, the initial overestimate for the number of biclusters was set to $K^* = 50$.

From \cite{Nicholls2022.12.07.519476}, we obtained a list of 56 genes associated with the IFN signature, 51 of which were found present in the preprocessed dataset. To focus on sparse, IFN-related biclusters, we filtered the results from both methods to include only biclusters with less than 50\% of the total number of samples and more than 6 IFN genes.

\begin{figure}[H]
  \centering
  \includegraphics[angle=90,width=0.6\textheight]{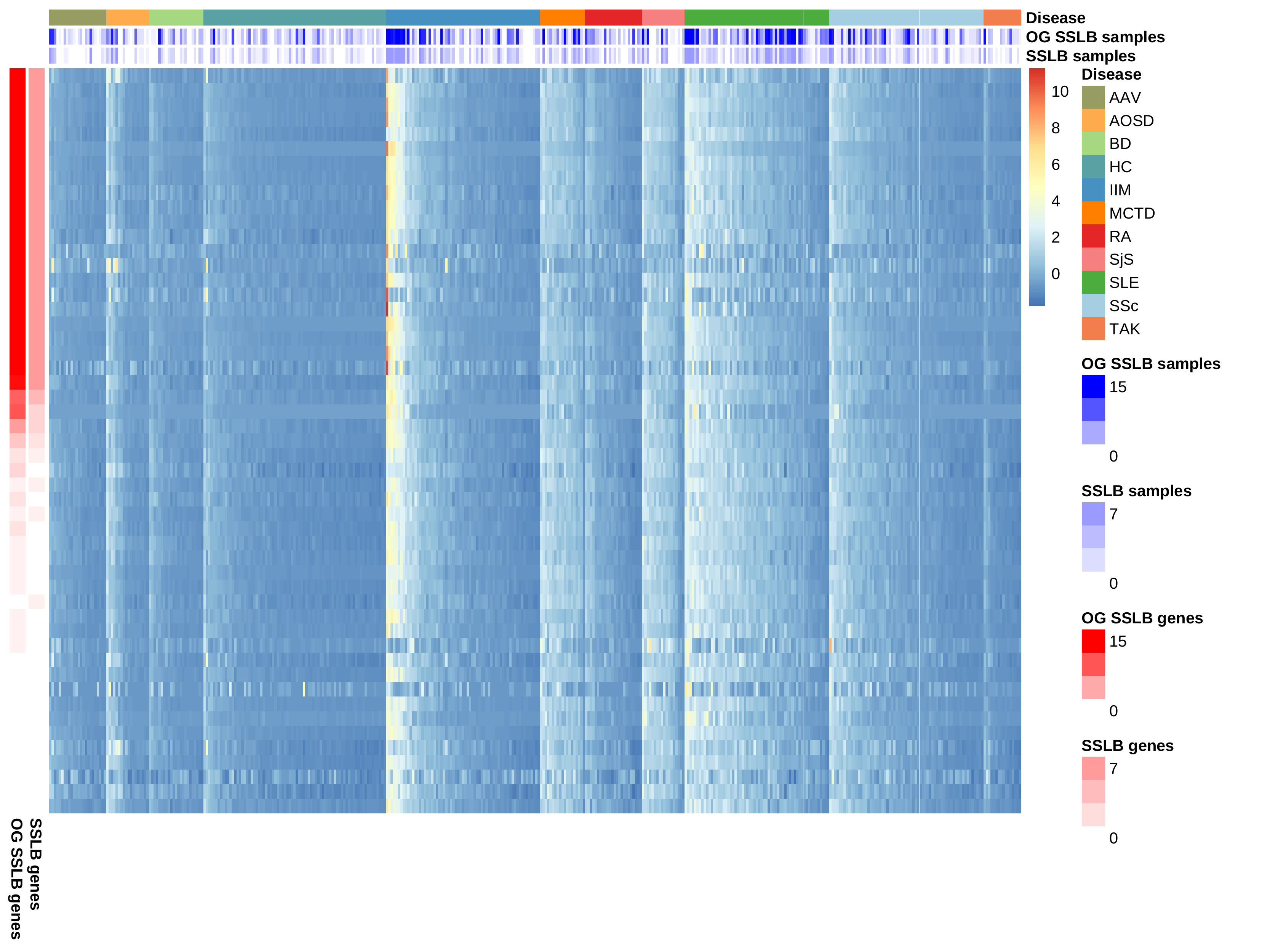}
  \caption{Heatmap of standardised gene expression data from the ImmuNexUT study, where each column represents a sample and each row corresponds to an interferon (IFN) signature gene. The data has been centred and scaled. Annotations indicate the frequency with which each gene or sample is detected by the SSLB and OG-SSLB algorithms. Samples are grouped by disease.}
  \label{fig:heatmap_immunex_ut}
\end{figure}

The results are first summarised in Figure \ref{fig:heatmap_immunex_ut}, where a heatmap of the standardised gene expression data is shown. As can be seen, SSLB and OG-SSLB generally identified the same samples and genes forming the IFN biclusters, but that the strength of OG-SSLB lay in its more frequent identification of such biclusters (18/20 compared to 7/20).

These results have also been summarised in Table \ref{tab:real_data_exp_median_patients_genes}. As can be seen, OG-SSLB is able to produce a greater number of replicates in which sparse IFN-related biclusters were detected, in comparison to SSLB.

\begin{table}[h]
\centering
\caption{Results from 20 replicates of applying SSLB and OG-SSLB to the ImmuNexUT real data, focusing on sparse IFN-related biclusters (\textless50\% of samples, \textgreater6 IFN genes).}\label{tab:real_data_exp_median_patients_genes}%
\begin{tabular}{lcc}
\toprule
Method & \begin{tabular}{@{}l} Replicates with at least one bicluster\\that meets sparse filtering cond.\end{tabular} & \begin{tabular}{@{}l} Median \% SLE patients in\\bicluster's replicates\end{tabular} 
\\
\midrule
SSLB & 7 & 36 \\
OG-SSLB & 18 & 43.5 \\
\bottomrule
\end{tabular}
\end{table}

\begin{figure}[!h]
\centering
\includegraphics[width = 0.75\linewidth]{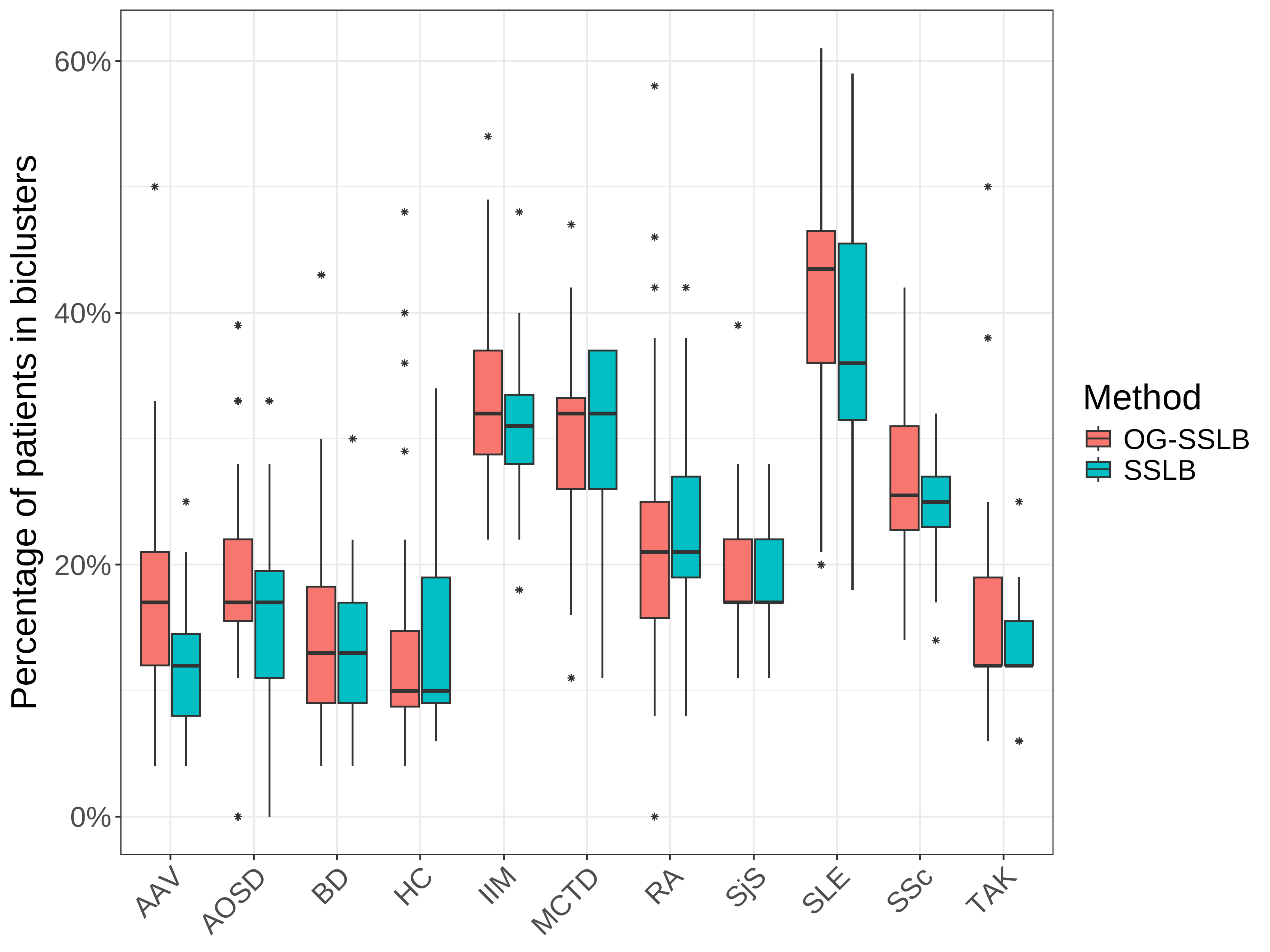}
\caption{Results from 20 replicates of applying SSLB and OG-SSLB to the ImmuNexUT real data, focusing on biclusters with less than 50\% of the total samples and more than 6 IFN genes: Distribution of the percentage of samples per disease.}\label{fig:real-data-results-perc-patients-per-dis}
\end{figure}

\begin{figure}[!h]
\centering
\includegraphics[width = 0.75\linewidth]{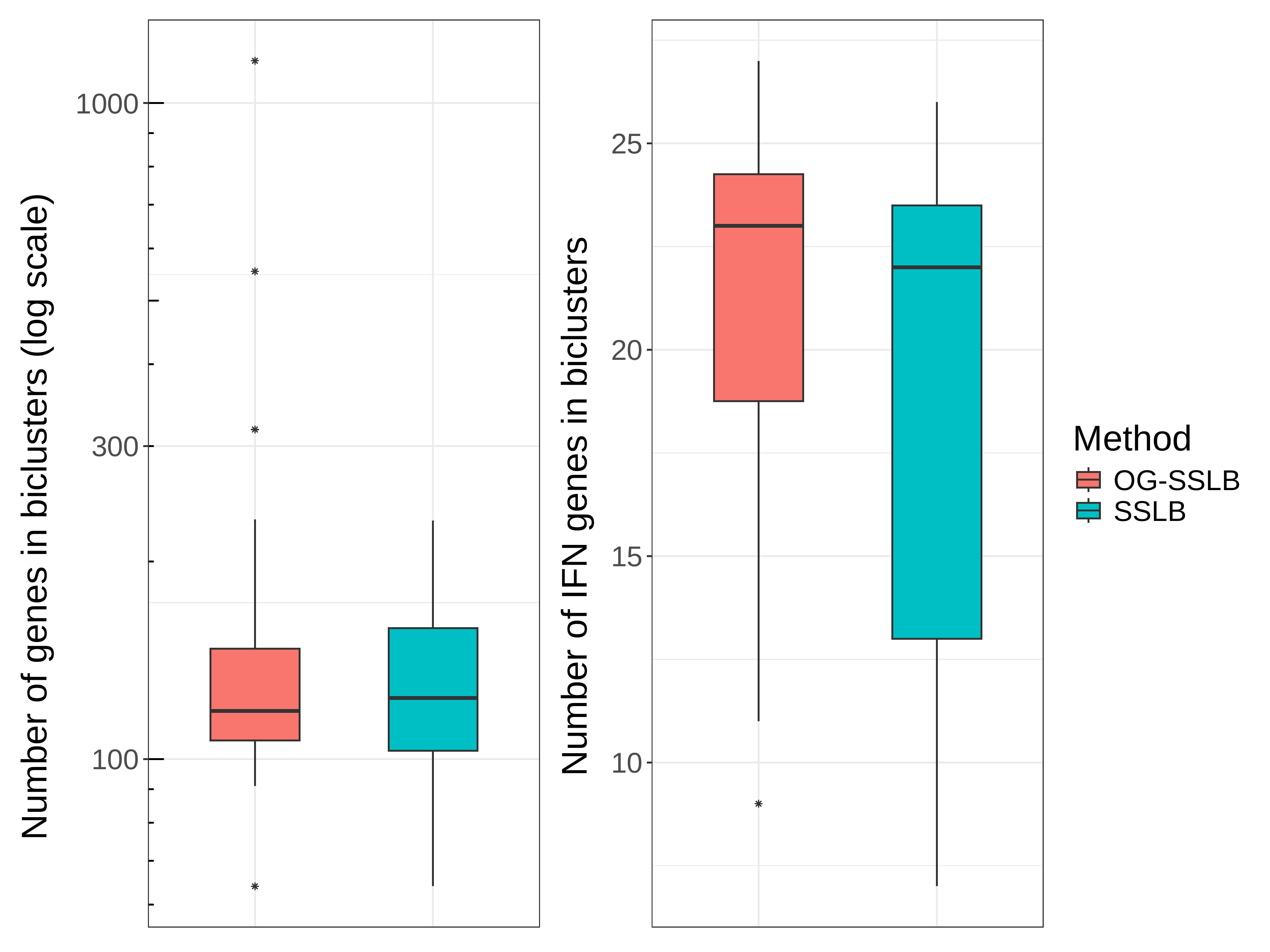}
\caption{Results from 20 replicates of applying SSLB and OG-SSLB to the ImmuNexUT real data, focusing on biclusters with less than 50\% of the total samples and more than 6 IFN genes. Left: Distribution of the total number of genes in biclusters identified by both methods. Right: Distribution of the number of IFN gene signatures in biclusters identified by both methods.}\label{fig:real-data-results-perc-genes-ifn}
\end{figure}

Furthermore, Figures \ref{fig:real-data-results-perc-patients-per-dis} and \ref{fig:real-data-results-perc-genes-ifn} show the distribution of samples and genes identified by the SSLB and OG-SSLB runs. OG-SSLB identified biclusters under the specified conditions exhibit, in distribution, a higher percentage of SLE patients and a higher number of IFN gene signatures. 
While SLE patients exhibited the highest fraction of patients in the IFN biclusters, IFN signatures have been found in other IMD, and both methods found a higher fraction of patients in IFN biclusters for IIM, MCTD, RA, SjS and SSc. Concerning the associated run times, the SSLB algorithm required about 1900 seconds to run, whereas the OG-SSLB algorithm took approximately 44700 seconds.

\section{Conclusions and Future work}
\label{sec:conclusions}
In conclusion, our proposed algorithm, OG-SSLB, exhibits superior performance compared to the SSLB approach in both numerical and real-data experiments, particularly in its ability to estimate the number of biclusters more accurately and achieve higher consensus scores. The flexibility of OG-SSLB, particularly through its multinomial modelling framework, allows it to accommodate more complex clustering structures than the commonly used binomial models. While this improvement entails significantly higher computational costs due to iterative processes such as AGD and ULA, the enhanced precision and modelling capacity of OG-SSLB make it a valuable contribution to biclustering methodologies.

Our subsequent analyses will expand the OG-SSLB framework to the ImmuNexUT dataset, investigating other cell types and detecting new gene expression signatures rather than focusing solely on predefined ones. We aim to identify similarities between diseases, facilitated by the bicluster overlapping allowed by this method.

Additionally, we will explore other machine learning alternatives to multinomial logistic regression, such as support vector machines and naive Bayes, which may offer more robust solutions for integrating disease information into the biclustering framework. An especially promising approach could be the introduction of deep learning classifiers, which may better capture the potential non-linearity of boundary classes, thereby further enhancing the quality of the incorporated disease information. While deep learning classifiers can identify more complex patterns in the data, they would require significantly larger computational resources compared to the multinomial logistic regression model. Nonetheless, the shift from multinomial logistic regression to deep learning or other machine learning approaches presents exciting opportunities to improve classification accuracy while addressing the challenges inherent in genomic data analysis.

The source code to reproduce the results in this paper is available online at \url{https://github.com/luisvargasmieles/OGSSLB-examples}.

\section*{Acknowledgements}
CW and LAVM are funded by the Wellcome Trust (WT220788). CW and PDWK are supported by the Medical Research Council (MC\textunderscore UU\textunderscore 00040/01 and MC\textunderscore UU\textunderscore 00040/05, respectively).

\section*{Data Availability}
The dataset for the Immune Cell Gene Expression Atlas (ImmuNexUT) experiment is in the public domain and available at the National Bioscience Database Center (NBDC), with the study accession code E-GEAD-397.

\section*{Statements and Declarations}
\subsection*{Competing Interests}
CW is a part-time employee of GSK. GSK had no role in this study or the decision to publish.

\newpage

\appendix
\section*{Appendix}

\section{Current SSLB model: E step}\label{sec:appendix_current_SSLB}
For completeness, we present the log posterior and the related expression for the E-step of SSLB. For additional details, refer to \cite{10.1214/20-AOAS1385}. The complete log posterior is as follows

\begin{align}
\log p(\boldsymbol{\Delta}, \boldsymbol{\Lambda}, \tilde{\mathbf{\Gamma}} \mid \mathbf{X} ) \propto & -\frac{1}{2} \sum_{i=1}^N\left\{\left(\mathbf{x}_{\mathbf{i}}-\mathbf{Z} \boldsymbol{\lambda}_{\boldsymbol{i}}\right)^T \mathbf{\Sigma}^{-1}\left(\mathbf{x}_{\mathbf{i}}-\mathbf{Z} \boldsymbol{\lambda}_{\boldsymbol{i}}\right)\right\}-\frac{N+\eta+2}{2} \sum_{j=1}^G \log \sigma_j^2 \nonumber \\ &-\sum_{j=1}^G \frac{\eta \xi}{2 \sigma_j^2} -\sum_{k=1}^{K^*} \log p \left(\mathbf{z}^{\mathbf{k}}\right)-\frac{1}{2} \sum_{i=1}^N \boldsymbol{\lambda}_{\mathbf{i}}^T \mathbf{D}_{\mathbf{i}} \boldsymbol{\lambda}_{\mathbf{i}}-\frac{1}{2} \sum_{i=1}^N \sum_{k=1}^{K^*} \log \tau_{i k} \nonumber \\
& -\frac{1}{2} \log \sum_{i=1}^N \sum_{k=1}^{K^*}\left[\left(1-\tilde{\gamma}_{i k}\right) \widetilde{\omega}_0 \exp \left(- \frac{\widetilde{\omega}_0 \tau_{i k}}{2}\right) \right.
 \left. +\tilde{\gamma}_{i k} \widetilde{\omega}_1 \exp \left(- \frac{\widetilde{\omega}_1 \tau_{i k}}{2} \right)\right] \nonumber \\
&+\sum_{k=1}^{K^*}\left[\left(\sum_{i=1}^N \tilde{\gamma}_{i k}\right) \log \prod_{l=1}^k \nu_l+\left(N-\sum_{i=1}^N \tilde{\gamma}_{i k}\right) \log \left(1-\prod_{l=1}^k \nu_l\right)\right] \nonumber \\ &+(\tilde{\alpha}-1) \sum_{k=1}^{K^*} \log \nu_k, \label{eqn:log_post_no_prof_reg}
\end{align}

where $\mathbf{D}_{\mathbf{i}}=\operatorname{diag}\left\{\tau_{i 1}^{-1}, \ldots, \tau_{i K^*}^{-1}\right\}$ and $\boldsymbol{\Gamma}$, $\boldsymbol{\theta}$ have been margined out to have $p (\mathbf{Z})$ (see \cite{doi:10.1080/01621459.2016.1260469} for details). This allows us to define

\begin{align}
Q(\boldsymbol{\Delta} \mid \boldsymbol{\Delta}^{(t)}) \propto & -\frac{1}{2} \sum_{i=1}^N \left\{ \left( \mathbf{x}_{i} - \mathbf{Z} \left\langle \boldsymbol{\lambda}_{i} \right\rangle \right)^T \boldsymbol{\Sigma}^{-1} \left( \mathbf{x}_{i} - \mathbf{Z} \left\langle \boldsymbol{\lambda}_{i} \right\rangle \right) + \operatorname{tr} \left[ \mathbf{Z}^T \boldsymbol{\Sigma}^{-1} \mathbf{Z} \left( \left\langle \boldsymbol{\lambda}_{i} \boldsymbol{\lambda}_{i}^T \right\rangle - \left\langle \boldsymbol{\lambda}_{i} \right\rangle \left\langle \boldsymbol{\lambda}_{i} \right\rangle^T \right) \right] \right\} \nonumber \\
& -\frac{N+\eta+2}{2} \sum_{j=1}^G \log \sigma_j^2 - \sum_{j=1}^G \frac{\eta \xi}{2 \sigma_j^2} - \sum_{k=1}^{K^*} \log p \left( \mathbf{z}^{k} \right) \nonumber \\
& -\frac{1}{2} \sum_{i=1}^N \left\{ \left\langle \boldsymbol{\lambda}_{i} \right\rangle^T \mathbf{D}_{i} \left\langle \boldsymbol{\lambda}_{i} \right\rangle + \operatorname{tr} \left[ \mathbf{D}_{i} \left( \left\langle \boldsymbol{\lambda}_{i} \boldsymbol{\lambda}_{i}^T \right\rangle - \left\langle \boldsymbol{\lambda}_{i} \right\rangle \left\langle \boldsymbol{\lambda}_{i} \right\rangle^T \right) \right] \right\} \nonumber \\
& -\frac{1}{2} \sum_{i=1}^N \sum_{k=1}^{K^*} \log \tau_{i k} -\frac{1}{2} \sum_{i=1}^N \sum_{k=1}^{K^*} \left[ \left(1 - \left\langle \tilde{\gamma}_{i k} \right\rangle \right) \widetilde{\omega}_0^2 + \left\langle \tilde{\gamma}_{i k} \right\rangle \widetilde{\omega}_1^2 \right] \tau_{i k} \nonumber \\
& +\sum_{k=1}^{K^*} \left[ \left( \sum_{i=1}^N \left\langle \tilde{\gamma}_{i k} \right\rangle \right) \log \prod_{l=1}^k v_l + \left( N - \sum_{i=1}^N \left\langle \tilde{\gamma}_{i k} \right\rangle \right) \log \left(1 - \prod_{l=1}^k v_l \right) \right] +(\tilde{\alpha} - 1) \sum_{k=1}^{K^*} \log v_k ,\label{eqn:exp_log_post_no_prof_reg}
\end{align}
where $\mathbb{E}_{\mathbf{\Lambda}, \widetilde{\Gamma} \mid \Delta^{(t)}, \mathbf{X}}[W] = \langle W\rangle$.

\section{SSLB model with profile regression: E step}\label{sec:appendix_new_SSLB}
The log posterior for the SSLB model using profile regression can be expressed as follows

\begin{align}
\log p(\boldsymbol{\Delta}, \boldsymbol{\Lambda}, \tilde{\mathbf{\Gamma}} \mid \mathbf{X}, \mathbf{Y} ) \propto & -\frac{1}{2} \sum_{i=1}^N\left\{\left(\mathbf{x}_{\mathbf{i}}-\mathbf{Z} \boldsymbol{\lambda}_{\boldsymbol{i}}\right)^T \mathbf{\Sigma}^{-1}\left(\mathbf{x}_{\mathbf{i}}-\mathbf{Z} \boldsymbol{\lambda}_{\boldsymbol{i}}\right)\right\} -\frac{N+\eta+2}{2} \sum_{j=1}^G \log \sigma_j^2-\sum_{j=1}^G \frac{\eta \xi}{2 \sigma_j^2} \nonumber \\
&-\sum_{k=1}^{K^*} \log p \left(\mathbf{z}^{\mathbf{k}}\right)-\frac{1}{2} \sum_{i=1}^N \boldsymbol{\lambda}_{\mathbf{i}}^T \mathbf{D}_{\mathbf{i}} \boldsymbol{\lambda}_{\mathbf{i}}-\frac{1}{2} \sum_{i=1}^N \sum_{k=1}^{K^*} \log \tau_{i k} \nonumber \\
&  -\frac{1}{2} \log \sum_{i=1}^N \sum_{k=1}^{K^*}\left[\left(1-\tilde{\gamma}_{i k}\right) \widetilde{\omega}_0 \exp \left(- \frac{\widetilde{\omega}_0 \tau_{i k}}{2}\right)  +\tilde{\gamma}_{i k} \widetilde{\omega}_1 \exp \left(- \frac{\widetilde{\omega}_1 \tau_{i k}}{2} \right)\right] \nonumber \\
&+\sum_{k=1}^{K^*}\left[\left(\sum_{i=1}^N \tilde{\gamma}_{i k}\right) \log \prod_{l=1}^k \nu_l +  \left(N-\sum_{i=1}^N \tilde{\gamma}_{i k}\right) \log \left(1-\prod_{l=1}^k \nu_l\right)\right] \nonumber \\ & + (\tilde{\alpha}-1) \sum_{k=1}^{K^*} \log \nu_k \nonumber \\
& -\sum_{i=1}^N \sum_{l=1}^C y_{i l} \mathbf{w}_l^T \widetilde{\gamma}_i^{\prime}+\sum_{i=1}^N \log \left[\sum_{l=1}^C \exp \left(\mathbf{w}_l^T \widetilde{\gamma}_i^{\prime}\right)\right] \nonumber \\ &+ \frac{1}{2} \lambda_{\text{W}} \lVert \mathbf{W} \rVert_{F}^2. \label{eqn:log_post_w_prof_reg}
\end{align}

Hence, the corresponding equation for the E-step is now defined as

\begin{align}
Q(\boldsymbol{\Delta} \mid \boldsymbol{\Delta}^{(t)}) \propto & \begin{multlined}[t] -\frac{1}{2} \sum_{i=1}^N\left\{\left(\mathbf{x}_{\mathbf{i}}-\mathbf{Z}\left\langle\lambda_{\boldsymbol{i}}\right\rangle\right)^T \boldsymbol{\Sigma}^{-1}\left(\mathbf{x}_{\mathbf{i}}-\mathbf{Z}\left\langle\boldsymbol{\lambda}_{\boldsymbol{i}}\right\rangle\right)
\right. \\ \left. + \operatorname{tr}\left[\mathbf{Z}^T \mathbf{\Sigma}^{-1} \mathbf{Z}\left(\left\langle\boldsymbol{\lambda}_{\boldsymbol{i}} \lambda_{\boldsymbol{i}}^T\right\rangle-\left\langle\boldsymbol{\lambda}_{\boldsymbol{i}}\right\rangle\left\langle\boldsymbol{\lambda}_{\boldsymbol{i}}\right\rangle^T\right)\right]\right\} \end{multlined} \nonumber \\
& -\frac{N+\eta+2}{2} \sum_{j=1}^G \log \sigma_j^2-\sum_{j=1}^G \frac{\eta \xi}{2 \sigma_j^2} -\sum_{k=1}^{K^*} \log p \left(\mathbf{z}^{\mathbf{k}}\right) \nonumber \\
& -\frac{1}{2} \sum_{i=1}^N\left\{\left\langle\boldsymbol{\lambda}_{\boldsymbol{i}}\right\rangle^T \mathbf{D}_{\mathbf{i}}\left\langle\lambda_{\boldsymbol{i}}\right\rangle+\operatorname{tr}\left[\mathbf{D}_{\mathbf{i}}\left(\left\langle\boldsymbol{\lambda}_{\boldsymbol{i}} \boldsymbol{\lambda}_{\boldsymbol{i}}^T\right\rangle-\left\langle\boldsymbol{\lambda}_{\boldsymbol{i}}\right\rangle\left\langle\boldsymbol{\lambda}_{\boldsymbol{i}}\right\rangle^T\right)\right]\right\} \nonumber \\ &- \frac{1}{2} \sum_{i=1}^N \sum_{k=1}^{K^*} \log \tau_{i k} -\frac{1}{2} \sum_{i=1}^N \sum_{k=1}^{K^*}\left[\left(1-\left\langle\tilde{\gamma}_{i k}\right\rangle\right) \widetilde{\omega}_0^2+\left\langle\tilde{\gamma}_{i k}\right\rangle \widetilde{\omega}_1^2\right] \tau_{i k} \nonumber \\
& \begin{multlined}[t] +\sum_{k=1}^{K^*}\left[\left(\sum_{i=1}^N\left\langle\tilde{\gamma}_{i k}\right\rangle\right) \log \prod_{l=1}^k v_l \right. \\ + \left. \left(N-\sum_{i=1}^N\left\langle\tilde{\gamma}_{i k}\right\rangle\right) \log \left(1-\prod_{l=1}^k v_l\right)\right] \end{multlined} \nonumber \\
& +(\tilde{\alpha}-1) \sum_{k=1}^{K^*} \log v_k  -\sum_{i=1}^N \sum_{l=1}^C y_{i l} \mathbf{w}_l^T\left\langle\widetilde{\boldsymbol{\gamma}}_i^{\prime}\right\rangle \nonumber \\ & + \sum_{i=1}^N \Biggl\langle \log \left[\sum_{l=1}^C \exp \left(\mathbf{w}_l^T \widetilde{\boldsymbol{\gamma}}_i^{\prime}\right)\right] \Biggr\rangle + \frac{1}{2} \lambda_{\text{W}} \lVert \mathbf{W} \rVert_{F}^2, \label{eqn:exp_log_post_w_prof_reg}
\end{align}

and the variables at which the new $Q(\boldsymbol{\Delta} \mid \boldsymbol{\Delta}^{(t)})$ will be maximised in the M step are now $\boldsymbol{\Delta}=\{\mathbf{Z}, \boldsymbol{\Sigma}, \mathbf{T}, \mathbf{W}, \nu\}$.

\bibliographystyle{abbrv}
\bibliography{main}  

\begin{thebibliography}{10}

\bibitem{NIPS2011_ad972f10}
A.~Armagan, M.~Clyde, and D.~Dunson.
\newblock Generalized beta mixtures of gaussians.
\newblock In J.~Shawe-Taylor, R.~Zemel, P.~Bartlett, F.~Pereira, and K.~Weinberger, editors, {\em Advances in Neural Information Processing Systems}, volume~24. Curran Associates, Inc., 2011.

\bibitem{https://doi.org/10.1002/wics.1270}
E.~Bair.
\newblock Semi-supervised clustering methods.
\newblock {\em WIREs Computational Statistics}, 5(5):349--361, 2013.

\bibitem{10.1371/journal.pbio.0020108}
E.~Bair and R.~Tibshirani.
\newblock Semi-supervised methods to predict patient survival from gene expression data.
\newblock {\em PLOS Biology}, 2(4), 04 2004.

\bibitem{https://doi.org/10.1002/sim.10119}
J.~Beall, H.~Li, B.~Martin-Harris, B.~Neelon, J.~Elm, E.~Graboyes, and E.~Hill.
\newblock Bayesian hierarchical profile regression for binary covariates.
\newblock {\em Statistics in Medicine}, 43(18):3432--3446, 2024.

\bibitem{bohning1992multinomial}
D.~B{\"o}hning.
\newblock Multinomial logistic regression algorithm.
\newblock {\em Annals of the institute of Statistical Mathematics}, 44:197--200, 1992.

\bibitem{10.1214/09-AOAS285}
H.~A. Chipman, E.~I. George, and R.~E. McCulloch.
\newblock {BART: Bayesian additive regression trees}.
\newblock {\em The Annals of Applied Statistics}, 4(1):266 -- 298, 2010.

\bibitem{7e6dbf96-e106-36fd-b898-4be31ae7ec6e}
A.~S. Dalalyan.
\newblock Theoretical guarantees for approximate sampling from smooth and log-concave densities.
\newblock {\em Journal of the Royal Statistical Society. Series B (Statistical Methodology)}, 79(3):651--676, 2017.

\bibitem{de2021efficient}
V.~De~Bortoli, A.~Durmus, M.~Pereyra, and A.~F. Vidal.
\newblock Efficient stochastic optimisation by unadjusted langevin monte carlo: Application to maximum marginal likelihood and empirical bayesian estimation.
\newblock {\em Statistics and Computing}, 31:1--18, 2021.

\bibitem{douc2014nonlinear}
R.~Douc, E.~Moulines, and D.~Stoffer.
\newblock {\em Nonlinear Time Series: Theory, Methods and Applications with R Examples}.
\newblock Chapman \& Hall/CRC Texts in Statistical Science. Taylor \& Francis, 2014.

\bibitem{10.1214/16-AAP1238}
A.~Durmus and {\'E}.~Moulines.
\newblock {Nonasymptotic convergence analysis for the unadjusted Langevin algorithm}.
\newblock {\em The Annals of Applied Probability}, 27(3):1551--1587, 2017.

\bibitem{10.1093/bib/bbs032}
K.~Eren, M.~Deveci, O.~K\"{u}\c{c}\"{u}ktun\c{c}, and U.~V. \c{C}ataly\"{u}rek.
\newblock {A comparative analysis of biclustering algorithms for gene expression data}.
\newblock {\em Briefings in Bioinformatics}, 14(3):279--292, 07 2012.

\bibitem{10.1371/journal.pcbi.1004791}
C.~Gao, I.~C. McDowell, S.~Zhao, C.~D. Brown, and B.~E. Engelhardt.
\newblock Context specific and differential gene co-expression networks via bayesian biclustering.
\newblock {\em PLOS Computational Biology}, 12(7):1--39, 07 2016.

\bibitem{NIPS2005_2ef35a8b}
Z.~Ghahramani and T.~Griffiths.
\newblock Infinite latent feature models and the indian buffet process.
\newblock In Y.~Weiss, B.~Sch\"{o}lkopf, and J.~Platt, editors, {\em Advances in Neural Information Processing Systems}, volume~18. MIT Press, 2005.

\bibitem{doi:10.1137/0802032}
O.~G{\"u}ler.
\newblock New proximal point algorithms for convex minimization.
\newblock {\em SIAM Journal on Optimization}, 2(4):649--664, 1992.

\bibitem{Hastie2009}
T.~Hastie, R.~Tibshirani, and J.~Friedman.
\newblock {\em Linear Methods for Classification}, pages 101--137.
\newblock Springer New York, New York, NY, 2009.

\bibitem{10.1093/bioinformatics/btq227}
S.~Hochreiter, U.~Bodenhofer, M.~Heusel, A.~Mayr, A.~Mitterecker, A.~Kasim, T.~Khamiakova, S.~Van~Sanden, D.~Lin, W.~Talloen, L.~Bijnens, H.~W.~H. Göhlmann, Z.~Shkedy, and D.-A. Clevert.
\newblock {FABIA: factor analysis for bicluster acquisition}.
\newblock {\em Bioinformatics}, 26(12):1520--1527, 2010.

\bibitem{1339264}
D.~Jiang, C.~Tang, and A.~Zhang.
\newblock Cluster analysis for gene expression data: a survey.
\newblock {\em IEEE Transactions on Knowledge and Data Engineering}, 16(11):1370--1386, 2004.

\bibitem{10.1093/bioinformatics/btq470}
D.~C. Koestler, C.~J. Marsit, B.~C. Christensen, M.~R. Karagas, R.~Bueno, D.~J. Sugarbaker, K.~T. Kelsey, and E.~A. Houseman.
\newblock Semi-supervised recursively partitioned mixture models for identifying cancer subtypes.
\newblock {\em Bioinformatics}, 26(20):2578--2585, 08 2010.

\bibitem{LEVITIN19661}
E.~Levitin and B.~Polyak.
\newblock Constrained minimization methods.
\newblock {\em USSR Computational Mathematics and Mathematical Physics}, 6(5):1--50, 1966.

\bibitem{JSSv064i07}
S.~Liverani, D.~I. Hastie, L.~Azizi, M.~Papathomas, and S.~Richardson.
\newblock {PReMiuM: An R Package for Profile Regression Mixture Models Using Dirichlet Processes}.
\newblock {\em Journal of Statistical Software}, 64(7):1–30, 2015.

\bibitem{love2014moderated}
M.~I. Love, W.~Huber, and S.~Anders.
\newblock Moderated estimation of fold change and dispersion for rna-seq data with deseq2.
\newblock {\em Genome Biology}, 15(12):1--21, 2014.

\bibitem{1324618}
S.~Madeira and A.~Oliveira.
\newblock {Biclustering algorithms for biological data analysis: a survey}.
\newblock {\em IEEE/ACM Transactions on Computational Biology and Bioinformatics}, 1(1):24--45, 2004.

\bibitem{10.1111/rssc.12536}
L.~Meng, D.~Avram, G.~Tseng, and Z.~Huo.
\newblock Outcome-guided sparse k-means for disease subtype discovery via integrating phenotypic data with high-dimensional transcriptomic data.
\newblock {\em Journal of the Royal Statistical Society Series C: Applied Statistics}, 71(2):352--375, 03 2022.

\bibitem{MESKO2011223}
B.~Mesko, S.~Poliska, and L.~Nagy.
\newblock Gene expression profiles in peripheral blood for the diagnosis of autoimmune diseases.
\newblock {\em Trends in Molecular Medicine}, 17(4):223--233, 2011.

\bibitem{10.1093/biostatistics/kxq013}
J.~Molitor, M.~Papathomas, M.~Jerrett, and S.~Richardson.
\newblock {Bayesian profile regression with an application to the National survey of children's health}.
\newblock {\em Biostatistics}, 11(3):484--498, 2010.

\bibitem{10.1214/20-AOAS1385}
G.~E. Moran, V.~Ročkov{\'a}, and E.~I. George.
\newblock {Spike-and-slab Lasso biclustering}.
\newblock {\em The Annals of Applied Statistics}, 15(1):148 -- 173, 2021.

\bibitem{1370862715914709505}
Y.~Nesterov.
\newblock A method for solving the convex programming problem with convergence rate o(1/k2), 1983.

\bibitem{Nicholls2022.12.07.519476}
K.~Nicholls, P.~D.~W. Kirk, and C.~Wallace.
\newblock Bayesian clustering with uncertain data.
\newblock {\em bioRxiv}, 2022.

\bibitem{10.1093/bib/bbab140}
K.~Nicholls and C.~Wallace.
\newblock {Comparison of sparse biclustering algorithms for gene expression datasets}.
\newblock {\em Briefings in Bioinformatics}, 22(6):1--16, 05 2021.

\bibitem{nikolakis2023restoration}
D.~Nikolakis, P.~Garantziotis, G.~Sentis, A.~Fanouriakis, G.~Bertsias, E.~Frangou, D.~Nikolopoulos, A.~Banos, and D.~T. Boumpas.
\newblock Restoration of aberrant gene expression of monocytes in systemic lupus erythematosus via a combined transcriptome-reversal and network-based drug repurposing strategy.
\newblock {\em BMC genomics}, 24(207), 2023.

\bibitem{ota2021dynamic}
M.~Ota, Y.~Nagafuchi, H.~Hatano, K.~Ishigaki, C.~Terao, Y.~Takeshima, H.~Yanaoka, S.~Kobayashi, M.~Okubo, H.~Shirai, et~al.
\newblock Dynamic landscape of immune cell-specific gene regulation in immune-mediated diseases.
\newblock {\em Cell}, 184(11):3006--3021, 2021.

\bibitem{padilha2017systematic}
V.~A. Padilha and R.~J. G.~B. Campello.
\newblock A systematic comparative evaluation of biclustering techniques.
\newblock {\em BMC Bioinformatics}, 18(55), 2017.

\bibitem{PEETERS2003651}
R.~Peeters.
\newblock {The maximum edge biclique problem is NP-complete}.
\newblock {\em Discrete Applied Mathematics}, 131(3):651--654, 2003.

\bibitem{doi:10.1126/science.abf1970}
R.~K. Perez, M.~G. Gordon, M.~Subramaniam, M.~C. Kim, G.~C. Hartoularos, S.~Targ, Y.~Sun, A.~Ogorodnikov, R.~Bueno, A.~Lu, et~al.
\newblock Single-cell rna-seq reveals cell type–specific molecular and genetic associations to lupus.
\newblock {\em Science}, 376(6589), 2022.

\bibitem{10.1093/bioinformatics/btl060}
A.~Prelić, S.~Bleuler, P.~Zimmermann, A.~Wille, P.~Bühlmann, W.~Gruissem, L.~Hennig, L.~Thiele, and E.~Zitzler.
\newblock {A systematic comparison and evaluation of biclustering methods for gene expression data}.
\newblock {\em Bioinformatics}, 22(9):1122--1129, 2006.

\bibitem{bj/1178291835}
G.~O. Roberts and R.~L. Tweedie.
\newblock {Exponential convergence of Langevin distributions and their discrete approximations}.
\newblock {\em Bernoulli}, 2(4):341 -- 363, 1996.

\bibitem{10.1093/bioinformatics/btp616}
M.~D. Robinson, D.~J. McCarthy, and G.~K. Smyth.
\newblock {edgeR: a Bioconductor package for differential expression analysis of digital gene expression data}.
\newblock {\em Bioinformatics}, 26(1):139--140, 2009.

\bibitem{10.1093/jrsssc/qlad097}
A.~Rouanet, R.~Johnson, M.~Strauss, S.~Richardson, B.~D. Tom, S.~R. White, and P.~D.~W. Kirk.
\newblock Bayesian profile regression for clustering analysis involving a longitudinal response and explanatory variables.
\newblock {\em Journal of the Royal Statistical Society Series C: Applied Statistics}, 73(2):314--339, 2023.

\bibitem{doi:10.1080/01621459.2016.1260469}
V.~Ročkov{\'a} and E.~I. George.
\newblock The spike-and-slab lasso.
\newblock {\em Journal of the American Statistical Association}, 113(521):431--444, 2018.

\bibitem{salzo2012inexact}
S.~Salzo and S.~Villa.
\newblock Inexact and accelerated proximal point algorithms.
\newblock {\em Journal of Convex analysis}, 19(4):1167--1192, 2012.

\bibitem{10.1093/bioinformatics/18.suppl_1.S136}
A.~Tanay, R.~Sharan, and R.~Shamir.
\newblock {Discovering statistically significant biclusters in gene expression data}.
\newblock {\em Bioinformatics}, 18:S136--S144, 2002.

\bibitem{pmlr-v2-teh07a}
Y.~W. Teh, D.~Grür, and Z.~Ghahramani.
\newblock Stick-breaking construction for the indian buffet process.
\newblock In {\em Proceedings of the Eleventh International Conference on Artificial Intelligence and Statistics}, volume~2 of {\em Proceedings of Machine Learning Research}, pages 556--563, San Juan, Puerto Rico, 2007. PMLR.

\bibitem{doi:10.1056/NEJMoa021967}
M.~J. van~de Vijver, Y.~D. He, L.~J. van~'t Veer, H.~Dai, A.~A. Hart, D.~W. Voskuil, S.~G. J., J.~L. Peterse, C.~Roberts, M.~J. Marton, M.~Parrish, D.~Atsma, A.~Witteveen, A.~Glas, L.~Delahaye, T.~van~der Velde, B.~H., R.~S., E.~T. Rutgers, S.~H. Friend, and R.~Bernards.
\newblock A gene-expression signature as a predictor of survival in breast cancer.
\newblock {\em New England Journal of Medicine}, 347(25):1999--2009, 2002.

\bibitem{doi:10.1137/20M1339829}
A.~F. Vidal, V.~De~Bortoli, M.~Pereyra, and A.~Durmus.
\newblock Maximum likelihood estimation of regularization parameters in high-dimensional inverse problems: An empirical bayesian approach part i: Methodology and experiments.
\newblock {\em SIAM Journal on Imaging Sciences}, 13(4):1945--1989, 2020.

\bibitem{10.1093/nargab/lqaa078}
Y.~Zhang, G.~Parmigiani, and W.~E. Johnson.
\newblock {ComBat-seq: batch effect adjustment for RNA-seq count data}.
\newblock {\em NAR Genomics and Bioinformatics}, 2(3), 2020.

\bibitem{Zhao2005}
Y.~Zhao and G.~Karypis.
\newblock Data clustering in life sciences.
\newblock {\em Molecular Biotechnology}, 31(1):55--80, 2005.

\end{thebibliography}

\end{document}